\documentclass[aps,prc,twocolumn,superscriptaddress,nofootinbib,floatfix]{revtex4-2}

\usepackage{amsmath,amssymb,bm}
\usepackage{graphicx}
\usepackage[colorlinks=true,citecolor=blue,linkcolor=blue,urlcolor=blue]{hyperref}

\newcommand{\LHe}{^{5}_{\Lambda}\mathrm{He}}
\newcommand{\LLi}{^{6}_{\Lambda}\mathrm{Li}}
\newcommand{\LHesix}{^{6}_{\Lambda}\mathrm{He}}
\newcommand{\kstar}{k^{*}}
\newcommand{\kstarsq}{k^{*2}}
\newcommand{\MeVc}{\,\mathrm{MeV}/c}
\newcommand{\GeVc}{\,\mathrm{GeV}/c}

\begin{document}

\title{Partial-wave decomposition of the $p$--$\LHe$ correlation function:\\
a femtoscopic route to the $\LLi$ resonance}

\author{Dongsheng~Li}
\affiliation{University of Science and Technology of China, Hefei, 230026 China}
\author{Yifei~Zhang}
\email{ephy@ustc.edu.cn}
\affiliation{University of Science and Technology of China, Hefei, 230026 China}

\date{\today}

\begin{abstract}
The hypernucleus $\LLi$ is unbound against proton emission, so its resonances appear in the
low-momentum $p$--$\LHe$ correlation function, which femtoscopy can measure. We compute this
correlation function in an $\alpha+\Lambda$ folding model of $\LHe$: phenomenological
$p\alpha$ and $p\Lambda$ interactions are averaged over the internal $\alpha$--$\Lambda$
motion, and a single $\Lambda N$ strength factor is fixed by the measured $\LHesix$ binding
energy. The model places the $\LLi$ $1^{-}$ state at $E_{R}=0.60^{+0.11}_{-0.10}$~MeV with
$\Gamma=0.18\pm0.05$~MeV and produces a pronounced peak in $C(\kstar)$ near $31\MeVc$. The
predicted peak position, $\kstar_{\rm peak}=\sqrt{2\mu E_{R}}=30.8\pm2.6\MeVc$, is uncertain
mainly because of the measured $B_\Lambda(\LHesix)$, not because of the $\Lambda N$
interaction. An exact partial-wave decomposition shows that the $P$ wave accounts for the entire enhancement over the Coulomb
baseline, $79\%$ of it from the $1^{-}$ channel. A Lednick\'y--Lyuboshitz analysis restricted
to $S$ waves therefore cannot describe this system. The maximum of the $1^{-}$ channel lies
within $1\%$ of $\kstar_{\rm peak}$, but the unresolved $2^{-}$ state and the momentum
resolution shift the measured maximum upward by $2.4\MeVc$ at $\sigma_{\kstar}=3\MeVc$. 

\end{abstract}

\maketitle

\section{Introduction}

The hyperon--nucleon ($YN$) interaction is one of the least constrained parts of the
baryon--baryon force. Because hyperons are short-lived, scattering experiments with hyperon
beams are difficult. The available cross sections are few, mostly integrated rather than
differential, and imprecise; no $\Lambda n$ data exist. Recent experiments have improved
the situation: $\Lambda p$ elastic scattering has been remeasured by
CLAS~\cite{CLAS2021Lp}, and $\Sigma^{\pm}p$ differential cross sections have been obtained at
J-PARC~\cite{JPARC2021Smp,JPARC2022Spp}. These data are still far less precise than $NN$
data, and they do not reach the near-threshold region that governs bound and nearly bound
systems. Hypernuclear binding energies therefore still carry most of the empirical
weight~\cite{Gal2016,Hashimoto2006}. The interaction also matters for dense matter: the
strengths of the $\Lambda N$ and $\Lambda NN$ forces determine whether hyperons appear in
neutron-star cores, and this ``hyperon puzzle'' depends on exactly the least constrained
components~\cite{Tolos2020,Lonardoni2015}.

In recent years, femtoscopy has emerged as a new source of information on the $YN$
interaction. Two-particle correlation functions measured at small relative momentum in
$pp$, $p$--A and A--A collisions probe the final-state interaction near threshold. The method
requires no hyperon beam, provides large pair samples, and extends to systems for which a
scattering experiment is impractical, such as hypernucleus--nucleon
pairs~\cite{Lisa2005,Fabbietti2021}. The $p\Lambda$ correlation has been measured with high
precision~\cite{ALICE2022pLambda}, the $p\Sigma^{+}$ correlation has been extracted for the
first time~\cite{ALICE2025pSigma}, and $d$--$\Lambda$ correlations have been observed in
Au+Au collisions at RHIC~\cite{STAR2025dLambda}. Calculations have also begun to extend the
approach to light nuclei, with predictions for deuteron--hyperon femtoscopy~\cite{dY2026}.

A natural next step is the correlation between a proton and a \emph{hypernucleus}. The
$p$--$\LHe$ pair is well suited for three reasons. First, $\LHe$ is the most tightly bound and
most compact $s$-shell hypernucleus ($B_\Lambda=3.12\pm0.02$~MeV). Its core is a spin-zero
$\alpha$ particle, so a cluster description is well justified and the $\Lambda$ carries the
entire spin of the hypernucleus. Second, the compound system $\LLi$ is unbound: it is
expected to lie a few hundred keV above the $p+\LHe$ threshold and decays by proton
emission~\cite{Oishi2018}. Its resonances therefore appear directly as enhancements of the
$p$--$\LHe$ correlation function. Third, a comparison with $p$--$\alpha$ at the same source
size isolates the effect of the added $\Lambda$, and the bound mirror hypernucleus $\LHesix$
independently constrains the $\Lambda N$ strength that acts in $p$--$\LHe$.

The $A=6$ hypernuclei have been studied theoretically for four decades. Motoba, Band\=o and
Ikeda described $\LHesix$ and $\LLi$ in a microscopic $\alpha+\chi+\Lambda$ three-cluster
model and obtained their spectra, electromagnetic transition rates and magnetic
moments~\cite{Motoba1983}. Hiyama \textit{et al.} performed accurate $\alpha+\Lambda+N$
three-body calculations with the Gaussian expansion method, treating all rearrangement
Jacobi coordinates on an equal footing. They reproduced the measured binding energy of
$\LHesix$ and the $(K^{-},\pi^{-})$ value for $\LLi$, and identified a three-layer structure
in $\LHesix$: an $\alpha$ core, a $\Lambda$ skin and a neutron halo~\cite{Hiyama1996}.
Filikhin \textit{et al.} solved the same $\alpha\Lambda n$ problem with configuration-space
Faddeev equations, found the $1^{-}$ state underbound, and suggested that this may indicate a
departure from a pure three-body cluster structure~\cite{Filikhin2014}. Oishi studied $\LLi$
as a one-proton emitter in a time-dependent three-body model. He showed that the emission
observables are sensitive to the $p\Lambda$ scattering length, and that adding the $\Lambda$
lowers both the energy and the width of the resonance relative to the $p\alpha$
case~\cite{Oishi2018}. In parallel, \emph{ab initio} methods have reached the $p$ shell. They
include the no-core shell model (NCSM) with chiral interactions and explicit
$\Lambda$--$\Sigma$ conversion~\cite{Wirth2014}, its extension to $YNN$ forces induced by the
similarity renormalization group (SRG)~\cite{WirthRoth2016,WirthRoth2018}, the Jacobi
NCSM~\cite{Le2020}, and NCSM calculations of charge-symmetry breaking (CSB) in the $A=7$ and
$8$ systems~\cite{Le2023}. Pionless effective field theory combined with the stochastic
variational method~\cite{Contessi2018}, and auxiliary-field diffusion Monte
Carlo~\cite{Lonardoni2014}, complete the picture. The \emph{ab initio} and
effective-field-theory studies consistently find that
two-body $YN$ forces alone overbind, and that repulsive $\Lambda NN$ or induced $YNN$ terms
are needed to restore agreement with experiment. Physically, these terms reflect the
suppression of $\Lambda$--$\Sigma$ conversion in the medium.

None of these calculations, however, yields the quantity that a femtoscopy experiment
measures. Bound-state methods do not address the continuum, and the Faddeev and
time-dependent treatments describe the resonance but not the correlation function. The
correlation function requires Coulomb-distorted scattering wave functions, summed over
partial waves and folded with an emission source in the Koonin--Pratt formula. This work
fills that gap.

We use a deliberately simple description: a two-body folding model in which $\LHe$ is a
frozen cluster. Core polarisation, explicit breakup and three-body forces are omitted; their
net effect is absorbed into a single strength factor, calibrated on the measured $\LHesix$
binding energy. In return, every ingredient of the correlation function is under explicit
control, and the signal can be decomposed exactly into its $(l,J)$ components. The central
result of this decomposition is that the observable is almost entirely a $P$-wave
phenomenon, which determines how a measurement should be analysed. Because no $p$--$\LHe$
correlation function has been measured yet, Sec.~\ref{sec:exp} also describes how such a
measurement can be made.

\section{Formalism}

\subsection{Correlation function and channel decomposition}

The correlation function is evaluated from the Koonin--Pratt
relation~\cite{Koonin1977,LL1982},
\begin{equation}
  C(\kstar)=\int d^{3}r\;S(r)\,\bigl|\Psi_{\kstar}(\bm r)\bigr|^{2},
  \label{eq:KP}
\end{equation}
where $\kstar$ is the relative momentum in the pair rest frame and $S(r)$ is a static
Gaussian source,
\begin{equation}
  S(r)=\bigl(4\pi R^{2}\bigr)^{-3/2}\exp\!\left(-\frac{r^{2}}{4R^{2}}\right).
\end{equation}
The proton and $\LHe$ are distinguishable at the cluster level, so no quantum-statistics
term appears. The Pauli principle between the proton and the $\alpha$ nucleons enters
through the $p\alpha$ interaction (Sec.~\ref{sec:input}).

Since the $\alpha$ core has $J^{\pi}=0^{+}$, the spin of $\LHe$ is that of the $\Lambda$,
and the pair spin $S=0,1$ coincides with the $p\Lambda$ singlet and triplet. Including the
proton orbital angular momentum $l$, we work in the basis of total angular momentum $J$.
Expanding $\Psi$ in partial waves and inserting it in Eq.~(\ref{eq:KP}) gives an
\emph{exactly additive} decomposition,
\begin{equation}
  C(\kstar)=C_{\rm Coul}(\kstar)+\sum_{l,J}\Delta C_{lJ}(\kstar),
  \label{eq:decomp}
\end{equation}
\begin{equation}
  \Delta C_{lJ}=\frac{2J+1}{4}\!\int\! 4\pi r^{2}S(r)\,
  \frac{\sum_{ab}\bigl|u^{+}_{ab}\bigr|^{2}-n_{lJ}F_{l}^{2}}{(\kstar r)^{2}}\,dr ,
  \label{eq:dC}
\end{equation}
where $F_{l}(\eta,\kstar r)$ is the regular Coulomb function, $u^{+}_{ab}$ are the
outgoing-wave-normalised radial solutions in channel $(l,J)$, $n_{lJ}$ is the number of
coupled $j_p=l\pm\tfrac12$ components, and $(2J+1)/4$ is the statistical weight. The
baseline
\begin{equation}
  C_{\rm Coul}(\kstar)=\int\! 4\pi r^{2}S(r)\sum_{l}(2l+1)\frac{F_{l}^{2}}{(\kstar r)^{2}}\,dr
\end{equation}
is the pure point-Coulomb result. Each $\Delta C_{lJ}$ is the correlation that one channel
adds to the Coulomb baseline, and each can be switched on and off independently;
Sec.~\ref{sec:pw} uses this decomposition throughout.

With $Z_{1}Z_{2}=2$ and $\mu=785.9$~MeV the Coulomb Bohr radius is $a_{B}=17.2$~fm. The
Coulomb repulsion is therefore strong: $C(\kstar)\to0$ as $\kstar\to0$, and the wave
functions must be built from Coulomb functions rather than plane waves. This alone makes
the analytic Lednick\'y--Lyuboshitz (LL) formula, which uses asymptotic wave functions and
$S$-wave parameters only, a poor starting point; Sec.~\ref{sec:ere} shows that the
partial-wave content makes it inapplicable. Albaladejo \textit{et al.}~\cite{Albaladejo2025}
have shown that extracting hadron--hadron amplitudes from charged-pair correlation
functions requires finite-range and off-shell corrections beyond the asymptotic amplitude.
We include these corrections exactly, by solving the radial equations instead of
parametrising their asymptotic form.

\subsection{Folding potential}
\label{sec:fold}

This paper treats the $A=6$ system as a nucleon moving in the field of a frozen $\LHe$. The full three-body solution of Hiyama \textit{et al.}~\cite{Hiyama1996} supports
this two-body reduction quantitatively. In the $1^{-}$ ground state of $\LHesix$, the
$\alpha+\Lambda$ subsystem is in its ground state, i.e.\ in $\LHe$, with probability
$95\%$; the $\LHe+N$ rearrangement channel alone, in a basis with $l,L\le1$, almost
reproduces the measured binding; and the $\alpha$--$\Lambda$ separation shrinks only from
$2.79$~fm in free $\LHe$ to $2.57$~fm in $\LHesix$. The $\LHe$ cluster thus survives the
addition of a nucleon essentially intact. A frozen core neglects a $5\%$ admixture and an
$8\%$ contraction of the core. Their residual effect is absorbed into the empirical
strength factor introduced in Sec.~\ref{sec:A6}, and Sec.~\ref{sec:limits} discusses what
this costs.

\begin{figure}[tb]
\centering
\includegraphics[width=\columnwidth]{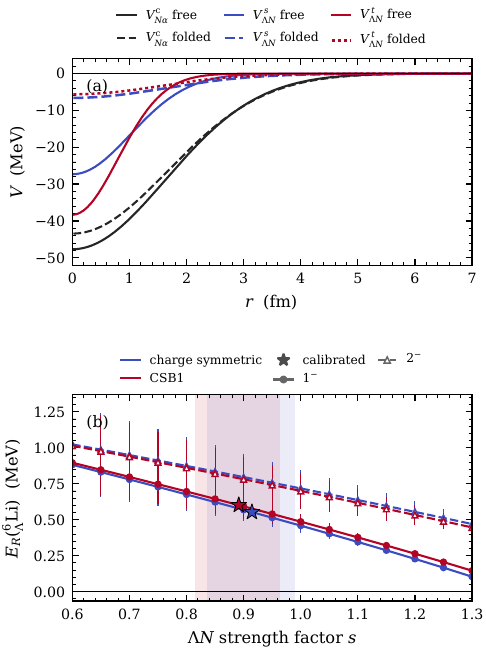}
\caption{(a) Two-body potentials before (solid) and after (broken) folding over the internal
$\alpha$--$\Lambda$ motion. The $\Lambda N$ singlet (blue) and triplet (red) Gaussians lose
roughly four fifths of their depth and gain about $1$~fm of range, because the $\Lambda$ is
spread over the halo of $\LHe$. The $N\alpha$ central term (black) barely changes, because
the $\alpha$ is compact. A single overall strength factor, rather than a renormalisation of
each term, is therefore the natural way to absorb what the frozen core leaves out.
(b) $\LLi$ resonance positions against the $\Lambda N$ strength factor $s$
(Sec.~\ref{sec:A6}), for charge symmetry (blue) and CSB1 (red); solid lines and filled circles
mark the $1^{-}$, dashed lines and open triangles the $2^{-}$. The bars show $\Gamma/2$, not an
uncertainty. The vertical bands are the ranges of $s$ that reproduce $B_\Lambda(\LHesix)$ in
each scenario, and the stars mark the $1^{-}$ prediction at each calibrated $s$.}
\label{fig:inputs}
\end{figure}

We describe $\LHe$ as a frozen $\alpha+\Lambda$ cluster with relative coordinate
$\bm\rho=\bm r_{\Lambda}-\bm r_{\alpha}$ and internal wave function $\phi(\rho)$. With $\bm r$
measured from the $\LHe$ centre of mass, the proton is at $\bm r_{p\alpha}=\bm
r+c_{\alpha}\bm\rho$ from the $\alpha$ and at $\bm r_{p\Lambda}=\bm r-c_{\Lambda}\bm\rho$
from the $\Lambda$, where $c_{\alpha}=m_{\Lambda}/M=0.23$, $c_{\Lambda}=m_{\alpha}/M=0.77$,
and $M$ is the mass of $\LHe$. Averaging both two-body interactions over the internal motion
gives
\begin{equation}
\begin{split}
  V(\bm r)=\int d^{3}\rho\,|\phi(\rho)|^{2}
  \Bigl[&V_{p\alpha}\bigl(\bm r+c_{\alpha}\bm\rho\bigr)\\
       &+V_{p\Lambda}\bigl(\bm r-c_{\Lambda}\bm\rho\bigr)\Bigr]+V_{C}(r).
\end{split}
  \label{eq:fold}
\end{equation}
Both terms must be folded, because the $\alpha$ is also displaced from the centre of mass.
Since $c_{\alpha}\ll c_{\Lambda}$, folding changes the $p\alpha$ depth by only about $10\%$
but smears the $p\Lambda$ term strongly. For Gaussian form factors the angular average is
analytic,
\begin{equation}
  \bigl\langle e^{-|\bm r+c\bm\rho|^{2}/b^{2}}\bigr\rangle_{\hat\rho}
  =e^{-(r-c\rho)^{2}/b^{2}}\,\frac{1-e^{-z}}{z},\qquad z=\frac{4rc\rho}{b^{2}},
\end{equation}
so Eq.~(\ref{eq:fold}) reduces to a one-dimensional quadrature over $\rho$.
Figure~\ref{fig:inputs}(a) shows the result. Folding conserves the volume integral (to one
part in $10^{6}$ numerically) but makes the potential shallower and longer-ranged: the
$p\Lambda$ Gaussians, $30$--$42$ MeV deep in free space, become $6$--$8$ MeV deep and
extend to $3$--$4$~fm. This dilution is the most important structural effect in the model.

The Coulomb term $V_{C}$ is that of a uniformly charged sphere of radius $R_{C}=2.36$~fm,
obtained by folding the $\alpha$ charge radius with the $\alpha$ centre-of-mass motion
inside $\LHe$; the corresponding $p\alpha$ value is $2.17$~fm. The nuclear part retains only
the difference from the point-Coulomb potential, so the phase shifts extracted below are
Coulomb-modified nuclear phase shifts in the standard sense.

Because the $\alpha$ has spin zero, the $p\Lambda$ spin--spin force acts directly as
$\bm\sigma_{p}\!\cdot\!\bm\sigma_{\Lambda}$ between the proton and $\LHe$. Together with the
$p\alpha$ spin--orbit force, which splits $P_{3/2}$ from $P_{1/2}$, it produces the
negative-parity multiplet: $J^{\pi}=2^{-}$ arises from $P_{3/2}$ alone, $0^{-}$ from
$P_{1/2}$ alone, and $1^{-}$ is a coupled $2\times2$ problem. The recoupling between the
$(l\,\tfrac12)j_p$ and $(\tfrac12\tfrac12)S$ schemes uses $6j$ symbols.

\subsection{Two-body input}
\label{sec:input}

All nuclear interactions are sums of Gaussians. Table~\ref{tab:input} summarises the input.

The $\Lambda\alpha$ interaction is the Isle potential of Kumagai-Fuse, Okabe, and
Akaishi~\cite{KumagaiFuse1995}, a two-range Gaussian with a repulsive core and a
longer-ranged attraction. Its original parameters give $B_\Lambda(\LHe)=3.10$~MeV; we
rescale it by a factor 1.0004 to reproduce 3.12~MeV. The resulting root-mean-square
$\alpha$--$\Lambda$ distance is 3.15~fm.

The $N\alpha$ interaction consists of a Gaussian central term and a Gaussian spin--orbit
term,
\begin{equation}
  V_{N\alpha}(r)=V_{0}e^{-r^{2}/b_{C}^{2}}
  +V_{LS}\,\bm l\!\cdot\!\bm s\;e^{-r^{2}/b_{LS}^{2}},
  \label{eq:Nalpha}
\end{equation}
with four parameters, of which we fit only two. The ranges are fixed \emph{a priori} at
$b_{C}=2.30$~fm and $b_{LS}=1.70$~fm. For comparison, the
Kanada--Kaneko--Nagata--Nomoto (KKNN) potential~\cite{KKNN1979} is a parity-dependent sum of
Gaussians with ranges between $0.6$ and $2.2$~fm; single Gaussians with the same volume
integral and mean-square radius as its odd-parity central and spin--orbit parts have ranges
of about $2.2$ and $1.6$~fm, close to ours. Each depth is adjusted to one observable:
\begin{itemize}
\item $V_{0}$ reproduces the $S$-wave $n\alpha$ scattering length $a_{S}=2.46$~fm. This is
  the free-nucleus value corresponding to the bound coherent scattering length $3.07(2)$~fm
  and the cross section $0.76(1)$~b of Ref.~\cite{Tilley2002}. We integrate the zero-energy
  radial equation with the Numerov solver of Sec.~\ref{sec:numerics}, read $a_{S}$ from the asymptotic form
  $u(r)\propto r-a_{S}$, and obtain $V_{0}=-47.69$~MeV by a one-dimensional root search.
  This is close to the $-47.6$~MeV depth of the odd-parity KKNN equivalent, which is the
  part that acts in the $P$ waves relevant here. Because the deep potential supports a
  Pauli-forbidden $0s$ state, $a_{S}(V_{0})$ has a pole where that state appears; the root
  search is restricted to the branch with exactly one forbidden state.
\item $V_{LS}$ reproduces the $^{5}$He $3/2^{-}$ resonance energy $E_{R}=0.80$~MeV, the
  evaluated $S$-matrix pole position of Ref.~\cite{Tilley2002} ($0.798$~MeV). In our model
  the speed plot of Sec.~\ref{sec:resextract} and the pole give the same energy ($0.800$
  and $0.798$~MeV), so the fit target is unambiguous on the model side. Fitting instead to
  the speed-plot value of the data phase shifts, $0.77$~MeV~\cite{CsotoHale1997}, changes
  the prediction negligibly (Sec.~\ref{sec:budget}). The root search gives
  $V_{LS}=-23.47$~MeV.
\end{itemize}
Both observables involve neutrons, so every proton quantity is a prediction. The model
places the $^{5}$Li $3/2^{-}$ resonance at $E_{R}=1.65$~MeV with $\Gamma=1.45$~MeV on the
speed plot, and its $S$-matrix pole at $1.645-i\,0.665$~MeV. Because $^{5}$Li is broad
($\Gamma/E_{R}\simeq0.8$), its ``resonance energy'' depends on the convention, and the
comparison must be made within one convention. Cs\'ot\'o and Hale~\cite{CsotoHale1997} list
$1.69$~MeV for the pole extracted from the data by an extended $R$-matrix analysis at
complex energies, $1.67$~MeV for the pole of their microscopic (RGM) model ($0.76$~MeV for
$^{5}$He), $1.76$ and $1.86$~MeV for conventional $R$-matrix fits to stripping and pickup
data, and $1.96$~MeV for the compilation value~\cite{AjzenbergSelove1988}. This spread of
$0.29$~MeV reflects definitions, not measurements. In the pole convention, our $^{5}$Li pole
lies $45$~keV below the data-derived pole ($1.645$ against $1.69$~MeV) and is $8\%$ wider
($1.33$ against $1.23$~MeV). For $^{5}$He the positions coincide ($0.798$~MeV) and our width
is again $8\%$ larger ($0.70$ against $0.65$~MeV~\cite{CsotoHale1997}). Our $^{5}$Li pole
agrees with the RGM pole to $25$~keV in position and $0.1\%$ in width. On the speed plot,
the data phase shifts give $0.77$~MeV for $^{5}$He and $1.53$~MeV for
$^{5}$Li~\cite{CsotoHale1997}, against our $0.80$ and $1.65$~MeV.

The $p\alpha$ $P_{1/2}$ resonance is a further test: it is not fitted, but depends on the
\emph{same} spin--orbit strength, with $\langle\bm l\!\cdot\!\bm s\rangle$ of opposite sign. Our pole lies at
$2.74-i\,2.92$~MeV, $0.44$~MeV below the data-derived pole at $3.18$~MeV~\cite{CsotoHale1997},
with an $11\%$ smaller width ($5.85$ against $6.60$~MeV); it agrees with the RGM pole
($2.70$, $6.25$~MeV) to $40$~keV and $6\%$. The model therefore reproduces the $3/2^{-}$
state, on which the $\LLi$ doublet is built, considerably better than the broad $1/2^{-}$.
Section~\ref{sec:sys-nalpha} quantifies the effect of the two fixed ranges. For the $S$ wave
we adopt the deep attractive representation with a Pauli-forbidden $0s$ state;
Sec.~\ref{sec:limits} discusses the phase-equivalent shallow repulsive alternative.

The $\Lambda N$ interaction is a spin-dependent Gaussian fitted to the singlet and triplet
scattering lengths and effective ranges of the chiral next-to-leading-order interaction with
a $600$~MeV cutoff, NLO19(600)~\cite{HMN2020}; all four are reproduced to better than
$0.1\%$. This interaction describes the ALICE $p\Lambda$ correlation
data~\cite{ALICE2022pLambda}. It is nevertheless the least certain ingredient of the
calculation, because the free-space interaction contains $\Lambda N$--$\Sigma N$ coupling
that is suppressed inside a nucleus. We therefore multiply the folded $\Lambda N$ term by a
single dimensionless strength factor $s$, determined from data in Sec.~\ref{sec:A6}.

The $\Lambda N$ input is constrained by $S$-wave data only, whereas the calculation retains
$p$--$\LHe$ partial waves up to $l=3$. This is consistent, because the $\Lambda N$ relative
angular momentum and the $p$--$\LHe$ orbital angular momentum are different quantum
numbers. After the folding of Eq.~(\ref{eq:fold}), the $\Lambda N$
force no longer appears as a $\Lambda N$ partial-wave interaction. It has been averaged over
the internal $\alpha$--$\Lambda$ motion and contributes a \emph{local, central,
spin-dependent potential between the proton and the $\LHe$ centre of mass}, which acts in
every $p$--$\LHe$ partial wave. The calculation does assume that the $\Lambda N$ interaction
can be represented by a local central force fitted to its $S$-wave scattering length and
effective range, i.e.\ that the $\Lambda N$ $P$-wave and tensor components are negligible.
The low momenta involved justify this: the $\Lambda$ inside $\LHe$ has a Fermi momentum of
only about $100\MeVc$, at which the centrifugal barrier suppresses $P$-wave $\Lambda N$
scattering. For the same reason, the folded potential retains essentially only the volume
integral and range of the $\Lambda N$ force, not its detailed shape.
Section~\ref{sec:sys-lamN} confirms quantitatively that, once the overall strength is
calibrated, the prediction is almost completely insensitive to the $\Lambda N$ form factor.

\begin{table*}[tb]
\caption{Two-body input and benchmark observables. Gaussian parameters are quoted as
$V(r)=V_{0}\exp(-r^{2}/b^{2})$. Quantities marked ``fitted'' define the parameters;
those marked ``predicted'' are outputs obtained without further adjustment.}
\label{tab:input}
\begin{ruledtabular}
\small
\begin{tabular}{llll}
Interaction & Parameters & Observable & Status\\
\colrule
$\Lambda\alpha$ (Isle~\cite{KumagaiFuse1995})
  & $V_{R}=450.4$~MeV, $b_{R}=1.25$~fm
  & $B_\Lambda(\LHe)=3.12$~MeV & fitted\\
  & $V_{A}=-404.9$~MeV, $b_{A}=1.41$~fm
  & $\sqrt{\langle r^{2}_{\alpha\Lambda}\rangle}=3.15$~fm & predicted\\[2pt]
$N\alpha$ central
  & $V_{0}=-47.69$~MeV, $b=2.30$~fm
  & $a_{S}(n\alpha)=2.46$~fm~\cite{Tilley2002} & fitted\\
$N\alpha$ spin--orbit
  & $V_{LS}=-23.47$~MeV, $b=1.70$~fm
  & $^{5}$He$(3/2^{-})$: $E_{R}=0.80$~MeV~\cite{Tilley2002} & fitted\\
  & & $^{5}$Li$(3/2^{-})$: $E_{R}=1.65$, $\Gamma=1.45$~MeV & predicted\\
  & & \quad ($S$-matrix pole $1.645-i\,0.665$; & \\
  & & \quad data pole~\cite{Tilley2002,CsotoHale1997}: $1.69$, $\Gamma=1.23$) & \\[2pt]
$\Lambda N$ singlet, NLO19(600)
  & $V_{0}=-31.85$~MeV, $b=1.434$~fm
  & $a_{s}=-2.91$~fm, $r_{s}=2.78$~fm & fitted\\
$\Lambda N$ triplet, NLO19(600)
  & $V_{0}=-42.08$~MeV, $b=1.126$~fm
  & $a_{t}=-1.41$~fm, $r_{t}=2.53$~fm & fitted\\
\end{tabular}
\end{ruledtabular}
\end{table*}

\subsection{Numerical procedure}
\label{sec:numerics}

For each $(l,J)$ the coupled radial equations
\begin{equation}
  u''_{ab}(r)=\Bigl[\tfrac{l(l+1)}{r^{2}}+\tfrac{2}{a_{B}r}-\kstarsq\Bigr]u_{ab}
  +\tfrac{2\mu}{\hbar^{2}}\sum_{c}V_{ac}(r)\,u_{cb}
\end{equation}
are integrated outward with a matrix Numerov scheme from $u\sim r^{l+1}$ and matched at
$r=14$~fm to Coulomb functions, which gives the $K$ matrix and the outgoing-wave solutions.
The CATS framework, widely used in femtoscopic analyses~\cite{CATS2018}, follows the same
strategy of solving the Schr\"odinger equation for the pair; here the strategy is extended to
coupled $(l,J)$ channels. Coulomb functions are generated on the radial grid from the exact power
series at small $\rho$, continued by Numerov integration; they agree with high-precision
reference values to $10^{-6}$ or better for all partial waves used. The point-Coulomb
baseline is summed to $l_{\max}=90$, and the nuclear interaction is retained for $l\le3$.
This is ample at the peak, where the $D$-wave contribution is $5\times10^{-4}$ and the $F$
wave $3\times10^{-6}$. The series converges more slowly at higher momentum, and we quote no
total $C(\kstar)$ above $100\MeVc$, where successive terms stand in the ratio
$D:F:G=0.112:0.0068:0.00041$. Bound states of the mirror system are obtained by sparse
shift-invert diagonalisation of the same coupled Hamiltonian and are converged to better
than 0.1~keV. As an internal check, the channel decomposition of Eq.~(\ref{eq:decomp})
reproduces the directly computed $C(\kstar)$ to a relative accuracy of $3\times10^{-16}$.

\subsection{Extraction of resonance parameters}
\label{sec:resextract}

Resonance positions and widths are obtained from the \emph{speed plot}. The Wigner--Smith
time delay of a scattering state is $\tau(E)=2\hbar\,d\delta/dE$, so a resonance appears as
a peak in the energy derivative of the (eigen)phase. For the Breit--Wigner form
\begin{equation}
  \frac{d\delta}{dE}=\frac{\Gamma/2}{(E-E_{R})^{2}+\Gamma^{2}/4},
  \label{eq:speed}
\end{equation}
the peak lies at $E_{R}$ and has height $2/\Gamma$. We compute $\delta(E)$ on a uniform
grid, form $d\delta/dE$ by finite differences, and locate the maximum by three-point
parabolic refinement; then $E_{R}=\mathrm{argmax}\,(d\delta/dE)$ and
$\Gamma=2/(d\delta/dE)_{\max}$. For coupled channels we use the eigenphase sum. The same
procedure is applied to the $^{5}$He and $^{5}$Li $P_{3/2}$ resonances and to the $\LLi$
multiplet.

A speed-plot $(E_{R},\Gamma)$ and an $S$-matrix pole are not the same pair of numbers. They
coincide only for a single pole without background, where Eq.~(\ref{eq:speed}) is exact;
any difference measures the non-resonant phase. We have therefore also located the poles.
Matching the interior solution to Coulomb functions gives $S=(1+iK)(1-iK)^{-1}$, and a pole
of $S$ is a zero of $\det[1-iK(E)]$ continued to complex $E$ on the second sheet,
$k=\sqrt{2\mu E}/\hbar c$ with $\mathrm{Re}\,k>0$ and $\mathrm{Im}\,k<0$. We find these zeros
by complex matrix Numerov integration, matched to Coulomb functions at complex $\eta$ and
$\rho$, to a residual $|\det[1-iK]|<3\times10^{-12}$.

Table~\ref{tab:resmethod} lists the results. For the narrow $\LLi$ states the pole positions
agree with the speed-plot positions to better than $0.3$~keV. The widths differ more: the
pole widths are $0.176$ and $0.310$~MeV, against speed-plot values of $0.178$ and
$0.320$~MeV. After the grid bias is removed (see below), these differences, $1.0\%$ for the
$1^{-}$ and $3.2\%$ for the $2^{-}$, are due to the non-resonant background. The effect is
larger for the broad $^{5}$Li resonance, where $\Gamma/E_{R}$ is three times larger: its pole
lies at $1.645-i\,0.665$~MeV, so the speed plot is high by $4.6$~keV in position and by
$9.3\%$ in width. For continuity with the hypernuclear literature we quote speed-plot values
throughout, except in the $A=5$ comparisons of Secs.~\ref{sec:input} and~\ref{sec:A6}, which
use the pole convention and are labelled accordingly. Pole widths are smaller by a few
percent, and no conclusion of this paper depends on the choice.

Table~\ref{tab:resmethod} includes two further determinations. The first fits the phase
over $E_{R}\pm\Gamma$ with a Breit--Wigner arctangent plus a linear background. It
reproduces the positions of the narrow states and, because the background term absorbs the
non-resonant phase, gives widths close to the pole widths. The second takes the energy at
which the phase passes $\pi/2$ modulo $\pi$. This criterion is systematically higher
($0.627$ versus $0.602$~MeV for the $\LLi$ $1^{-}$, and $2.25$ versus $1.65$~MeV for
$^{5}$Li), because it does not separate the resonant phase from the background; we do not
use it.

To check convergence we varied each numerical parameter about its production value.
Halving or doubling the radial step ($h=0.02$~fm), and moving the matching radius from $14$
to $12$ or $18$~fm, change neither $E_{R}$ nor $\Gamma$ by more than $0.01$~keV. (Extending
the outer end of the radial grid is not an independent test, because the outward
integration is read only at the matching radius.) The energy step $dE$ of the finite
differences requires more care. A central difference averages the speed function over a
window of $2\,dE$ and so overestimates $\Gamma$; for a Breit--Wigner the bias is
$\Gamma_{\rm meas}=\Gamma[1+4\,dE^{2}/3\Gamma^{2}]$, which we remove analytically. Without
this correction the $1^{-}$ width is $0.1784$, $0.1790$, $0.1812$ and $0.1961$~MeV at
$dE=0.005$, $0.01$, $0.02$ and $0.05$~MeV, a spread of $10\%$; with it, the width is
$0.1782$, $0.1782$, $0.1783$ and $0.1773$~MeV, a spread of $0.6\%$, and only $0.03\%$ over
the first three. We use $dE=0.01$~MeV and assign a $1\%$ numerical uncertainty to the
widths and none to the positions. States narrower than about $0.03$~MeV, which occur only
near threshold in the strength scans, are evaluated on a local grid with $dE=\Gamma/50$ and
checked against the pole.

\begin{table*}[tb]
\caption{Resonance parameters from the four determinations discussed in the text, in MeV.
``Speed'' is the method used throughout the paper; ``pole'' is the zero of
$\det[1-iK]$ on the second sheet. Values for the $\LLi$ states are at the calibrated
CSB1 strength.}
\label{tab:resmethod}
\begin{ruledtabular}
\small
\begin{tabular}{lccccccc}
 & \multicolumn{2}{c}{speed plot} & \multicolumn{2}{c}{$S$-matrix pole}
 & \multicolumn{2}{c}{Breit--Wigner fit} & $\delta=\pi/2$\\
State & $E_{R}$ & $\Gamma$ & $E_{R}$ & $\Gamma$ & $E_{R}$ & $\Gamma$ & $E_{R}$\\
\colrule
$^{5}$Li $3/2^{-}$ & 1.650 & 1.455 & 1.645 & 1.331 & 1.668 & 1.325 & 2.245\\
$\LLi$ $1^{-}$     & 0.602 & 0.178 & 0.602 & 0.176 & 0.602 & 0.176 & 0.627\\
$\LLi$ $2^{-}$     & 0.788 & 0.320 & 0.788 & 0.310 & 0.789 & 0.310 & 0.853\\
\end{tabular}
\end{ruledtabular}
\end{table*}

\section{Calibration on the $A=6$ mirror pair}
\label{sec:A6}

Adding a nucleon to $\LHe$ gives $\LLi$ (proton, unbound) and $\LHesix$ (neutron, bound). The
two are mirror systems: with a charge-symmetric $\Lambda N$ interaction, their potentials
differ only by the Coulomb term.
This allows a well-measured bound-state energy to calibrate a continuum prediction.

The $\Lambda$ separation energy of $\LHesix$ is measured relative to $^{5}$He$+\Lambda$,
\begin{equation}
  B_\Lambda(\LHesix)=E(^{5}\mathrm{He})+B_\Lambda(\LHe)+B_{n}.
\end{equation}
The compiled emulsion value $B_\Lambda(\LHesix)=4.18\pm0.10$~MeV from $31$
events~\cite{DavisPniewski1986,Gal2016} therefore corresponds to a neutron separation energy
$B_{n}=0.26\pm0.10$~MeV from $n+\LHe$, with $E(^{5}\mathrm{He})=0.80$~MeV as in the model.
The mass-table value $E(^{5}\mathrm{He})=0.735$~MeV~\cite{AME2020} would instead give
$B_{n}=0.32$~MeV and increase the fitted strength by $0.05$. The underlying emulsion data
are not fully consistent: the $17$ events of Ref.~\cite{Juric1973} alone give
$4.42\pm0.13$~MeV, and that paper's average over the $31$ events then available was
$4.25\pm0.10$~MeV. This spread indicates how well the binding energy is actually known.

\begin{figure}[tb]
\centering
\includegraphics[width=\columnwidth]{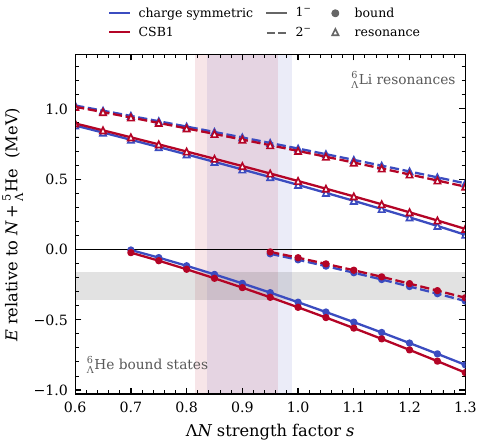}
\caption{The $A=6$ mirror pair as a function of the $\Lambda N$ strength factor $s$, with
the rest of the model fixed. Colour gives the $\Lambda N$ scenario (blue, charge symmetric;
red, CSB1), line style the spin (solid, $1^{-}$; dashed, $2^{-}$), and the marker the type
of state: filled circles are $\LHesix$ bound states, shown from the strength at which each
becomes bound, and open triangles are $\LLi$ resonance positions. All energies are measured
from the $N+\LHe$ threshold. The horizontal grey band is the measured
$B_\Lambda(\LHesix)=4.18\pm0.10$~MeV, and the vertical bands are the ranges of $s$ that
reproduce it in each scenario. The asymmetry between the mirror systems comes from the
Coulomb interaction and, for CSB1, from the $\Lambda p$--$\Lambda n$ difference; no
parameter is adjusted between them.}
\label{fig:mirror}
\end{figure}

Table~\ref{tab:A6} and Fig.~\ref{fig:mirror} show how the multiplet evolves with the
$\Lambda N$ strength factor $s$. Without a $\Lambda N$ interaction, both $A=6$ systems are
unbound, and the $\LLi$ $1^-$ and $2^-$ states lie near the $^{5}$Li-like energy of
$1.44$--$1.45$~MeV. As the interaction is switched on, the states move down rapidly and
become much narrower. At the free NLO19(600) strength, $\LHesix(1^-)$ is bound by
$0.375$~MeV, i.e.\ $B_\Lambda(\LHesix)=4.30$~MeV, slightly more than the measured
$4.18\pm0.10$~MeV. Reproducing the measured value requires
\begin{equation}
  s=0.915^{+0.075}_{-0.078}.
\end{equation}
This reduction is consistent with the expected in-medium suppression of
$\Lambda N$--$\Sigma N$ coupling and with the repulsive $\Lambda NN$ or induced $YNN$ terms
found in microscopic calculations~\cite{WirthRoth2016,Lonardoni2014,Contessi2018}. A full
three-body treatment would add binding that our frozen core lacks, so $s=0.915$ is expected to
be an upper estimate.

At the calibrated strength, the ground state of $\LHesix$ is $1^-$. The neutron occupies
the $P_{3/2}$ orbit with probability $99.6\%$, and the root-mean-square $n$--$\LHe$
separation is $5.54$~fm; combined with the $\alpha$--$\Lambda$ distance of the frozen core,
this gives $\bar{r}_{\alpha n}=5.59$~fm and $\bar{r}_{\Lambda n}=6.05$~fm. These are the
neutron halo and $\Lambda$ skin found in the three-body calculation of Hiyama
\textit{et al.}~\cite{Hiyama1996}, who obtain $\bar{r}_{\alpha n}=5.22$~fm and
$\bar{r}_{\Lambda n}=5.54$~fm. Our distances are $7$--$9\%$ larger because the frozen core
cannot contract: in their calculation the $\alpha$--$\Lambda$ separation shrinks from
$2.79$~fm in free $\LHe$ to $2.57$~fm in $\LHesix$. The $2^{-}$ state lies essentially at
threshold, and the model cannot decide whether it is bound; Hiyama \textit{et al.} place it
$0.11$~MeV \emph{above} the $n+\LHe$ threshold.

The corresponding mirror prediction is
\begin{equation}
  \LLi(1^{-}):\;E_{R}=0.553\;\mathrm{MeV},\quad \Gamma=0.149\;\mathrm{MeV},
\end{equation}
with the $2^{-}$ at $E_{R}=0.785$~MeV, $\Gamma=0.317$~MeV. The $1^{-}$ position agrees well
with the only previous calculation that quotes it: from the full $\alpha+\Lambda+N$
three-body problem, Hiyama \textit{et al.}~\cite{Hiyama1996} obtain $E_{R}=0.58$~MeV,
against our $0.55$~MeV ($0.60$~MeV with charge-symmetry breaking, Sec.~\ref{sec:csb}). Their
$2^{-}$ lies at $0.97$~MeV, against our $0.79$~MeV.

The rest of the paper uses two versions of the $\Lambda N$ interaction. The first is charge
symmetric (CS): the $\Lambda p$ and $\Lambda n$ interactions are identical. The results of
this section, including the strength scans, use this version. The second, CSB1,
includes the charge-symmetry breaking inferred from the $A=4$ mirror hypernuclei, so the
$\Lambda p$ scattering lengths differ from the $\Lambda n$ ones (Sec.~\ref{sec:csb}). The
calibration on $\LHesix$ constrains the $\Lambda n$ interaction, whereas the $p$--$\LHe$
correlation depends on the $\Lambda p$ interaction; CSB1 is therefore the more consistent
choice, and we quote it as the prediction. It is used in the uncertainty budget
(Sec.~\ref{sec:budget}), in the survey of chiral interactions (Sec.~\ref{sec:sys-chiral}),
and for the correlation functions from Sec.~\ref{sec:shape} onwards. The two versions differ
by $49$~keV in $E_{R}$ and by $1.3\MeVc$ in $\kstar_{\rm peak}$, well inside the uncertainty
of Sec.~\ref{sec:budget}. Every number below that depends on the choice is labelled.

Experiment constrains this prediction only loosely. Because $\LLi$ is unbound against proton
emission, it decays strongly long before it could decay weakly. It therefore cannot be
identified in emulsion, the technique that measured the binding energies of the other light
hypernuclei, and an emulsion search found no candidate~\cite{Juric1973}. The only
measurement of $B_\Lambda(\LLi)$ comes from the $^{6}$Li$(K^{-},\pi^{-})$ experiment of the
Heidelberg--Saclay--Strasbourg collaboration at the CERN Proton
Synchrotron~\cite{Bertini1981}. A $790\MeVc$ kaon beam converted neutrons of the target into
$\Lambda$ hyperons; the forward-going pions were momentum-analysed in a magnetic
spectrometer, and the mass difference between the hypernuclear system and the $^{6}$Li
target was reconstructed from the kaon and pion momenta. This mass difference equals
$m_\Lambda-m_n+B_n-B_\Lambda$, where $B_n$ is the binding energy of the converted neutron,
so $B_\Lambda$ follows only once $B_n$ is known. For the ground state the authors took the
valence-neutron separation energy of $^{6}$Li from the mass tables, $B_n=5.66$~MeV. The only
ground-state candidate is a weak transition at $177.2$~MeV, which the authors noted may
correspond to the $\LLi$ $1^{-}$ state; this transition gives $B_\Lambda=4.5\pm0.5$~MeV. The transition is
weak because forward-angle $(K^{-},\pi^{-})$ reactions transfer little momentum and
therefore favour $\Delta L=0$, in which the $\Lambda$ takes over the orbit of the converted
neutron. The $1^{-}$ ground state, with the $\Lambda$ in the $s$ orbit and the neutron hole
in the $p$ shell, requires $\Delta L=1$.

The same spectrum provides the only internal check of this value. A strong, narrow peak at
$195.5$~MeV corresponds to the conversion of a $1s$ neutron. Assuming that the $\Lambda$ is
bound as strongly to the $1s$-hole core as to the ground-state core, the authors deduced a
$1s$-neutron binding energy of $24.0\pm0.5$~MeV. This is $1.7$~MeV larger than the
$22.3$~MeV measured in $^{6}$Li$(p,d)$~\cite{Fagerstrom1976}, and they suggested that the
difference may be due to that assumption. The deduced binding energy depends only on the
separation between the two peaks, so the same discrepancy would arise if the weak
ground-state transition were displaced by $1.7$~MeV. The check therefore does not confirm
the ground-state assignment either.

With $B_\Lambda(\LHe)=3.12\pm0.02$~MeV~\cite{Juric1973} and the $^{5}$Li $3/2^{-}$ energy,
$B_\Lambda(\LLi)=4.5$~MeV places the $\LLi$ $1^{-}$ $0.3$--$0.6$~MeV above the $p+\LHe$
threshold. The range reflects the choice of $^{5}$Li energy: $1.69$~MeV in the current
evaluation~\cite{Tilley2002}, $1.96$~MeV in the older compilation~\cite{AjzenbergSelove1988}
and $1.97$~MeV in Ref.~\cite{Hiyama1996}. The older values reproduce the $600$~keV quoted in
Ref.~\cite{Bertini1981}. The $\pm0.5$~MeV uncertainty of the measurement adds to this range.
Our $0.60^{+0.11}_{-0.10}$~MeV lies within it, but the range is too wide to confirm or rule
out the prediction. The estimate $B_{p}(\LLi)\simeq-0.6$~MeV from separation-energy
systematics~\cite{Oishi2018} rests on the same sparse data; with the current $^{5}$Li energy
it corresponds to $B_\Lambda(\LLi)\approx4.2$~MeV, and Ref.~\cite{Oishi2018} assigns it an
uncertainty of $0.1$--$1$~MeV.

The width is even less constrained. Oishi~\cite{Oishi2018}, in a time-dependent three-body
treatment of proton emission, estimates the width of the $1^{-}$ to be $0.01$--$0.1$~MeV,
depending on the $p\Lambda$ scattering length. Our $\Gamma=0.15$--$0.18$~MeV lies above
this range: $1.5$--$2$ times his upper value and more than ten times his lower one. Our width
is itself uncertain. The $\pm0.10$~MeV on $B_\Lambda(\LHesix)$ alone gives
$\Gamma=0.178^{+0.053}_{-0.045}$~MeV in the default CSB1 scenario and
$0.149^{+0.053}_{-0.044}$~MeV with charge symmetry (Sec.~\ref{sec:budget}); the lower edges,
$0.13$ and $0.10$~MeV, approach or reach the upper end of his range. Because the width
decides whether the resonance can be resolved in a correlation measurement at all, this
difference is worth settling.

Across the whole range of strengths, the energy difference between mirror states is
$0.79$--$0.89$~MeV, and $0.81$~MeV at the calibrated point. This is essentially the
$^{5}$Li--$^{5}$He Coulomb displacement energy of the model, growing slowly with binding as
expected. In the pole convention, the model gives $0.847$~MeV for $A=5$, against $0.89$~MeV
from the data-derived poles~\cite{Tilley2002,CsotoHale1997} and $0.91$~MeV from the RGM
poles of Ref.~\cite{CsotoHale1997}. The $45$~keV deficit comes entirely from $^{5}$Li,
because the $^{5}$He pole is fitted, and it is much smaller than the $0.29$~MeV spread of
the quoted $^{5}$Li energies (Sec.~\ref{sec:input}).

The pole convention is appropriate for this comparison. The Coulomb shift acts on the pole
position, and for the narrow $\LLi$ state the pole and speed-plot positions
agree to $0.1$~keV (Table~\ref{tab:resmethod}). The speed-plot positions of the broad $A=5$
states, in contrast, also include the different penetrabilities of the neutron and proton
channels; on the speed plot the same comparison gives a $90$~keV \emph{excess}
($1.65-0.80$ against $1.53-0.77$~MeV~\cite{CsotoHale1997}). Since the prediction is this
displacement energy minus the measured $B_{n}(\LHesix)$, the deficit in the pole convention
biases our $E_{R}$ low by about $45$~keV. The frozen core points in the same direction, so we
regard $0.60$~MeV as, if anything, a lower estimate. The deficit is not caused by the
Coulomb radius: varying $R_{C}$ from $1.9$ to $2.8$~fm moves the displacement energy only
between $0.857$ and $0.819$~MeV, so even the smallest radius recovers only $10$ of the
$45$~keV.

\begin{table*}[tb]
\caption{Energies of the $A=6$ multiplet relative to the $N+\LHe$ threshold as a function of
the $\Lambda N$ strength factor $s$, for the charge-symmetric (CS) $\Lambda N$ interaction.
Negative entries are bound states; positive entries are resonances, with the width in
parentheses. All values in MeV. The row $s=0.915$ reproduces
$B_\Lambda(\LHesix)=4.18\pm0.10$~MeV; the CSB1 calibration is $s=0.891$
(Sec.~\ref{sec:csb}).}
\label{tab:A6}
\begin{ruledtabular}
\small
\begin{tabular}{lcccc}
 & \multicolumn{2}{c}{$\LHesix$} & \multicolumn{2}{c}{$\LLi$}\\

$s$ & $1^{-}$ & $2^{-}$ & $1^{-}$ & $2^{-}$\\
\colrule
$0$ (no $\Lambda N$) & $+0.626\,(0.500)$ & $+0.624\,(0.527)$ & $+1.449\,(1.048)$ & $+1.439\,(1.170)$\\
$0.6$ & $+0.092\,(0.026)$ & $+0.225\,(0.102)$ & $+0.882\,(0.395)$ & $+1.024\,(0.558)$\\
$0.8$ & $-0.116$ & $+0.082\,(0.022)$ & $+0.676\,(0.228)$ & $+0.874\,(0.398)$\\
$\mathbf{0.915}$ & $\mathbf{-0.260}$ & $-0.002$ & $\mathbf{+0.553\,(0.149)}$ & $+0.785\,(0.317)$\\
$1.0$ (free NLO19(600)) & $-0.375$ & $-0.074$ & $+0.459\,(0.099)$ & $+0.718\,(0.261)$\\
$1.2$ & $-0.666$ & $-0.263$ & $+0.228\,(0.018)$ & $+0.555\,(0.149)$\\
\end{tabular}
\end{ruledtabular}
\end{table*}

\section{The $p$--$\LHe$ correlation function}

\subsection{Overall shape and source dependence}
\label{sec:shape}

\begin{figure*}[tb]
\centering
\includegraphics[width=\textwidth]{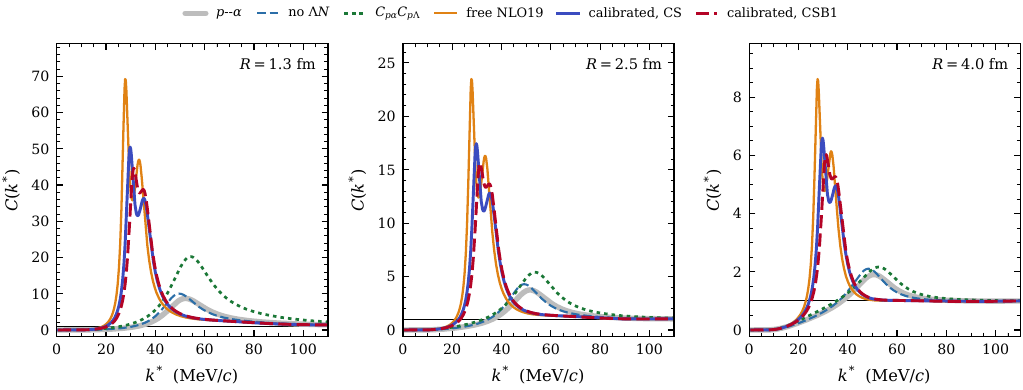}
\caption{$p$--$\LHe$ correlation functions for three source radii, without momentum
resolution (see Sec.~\ref{sec:exp-res} for its effect). Thick pale grey: $p$--$\alpha$, with the
$^{5}$Li $3/2^{-}$ peak near $52\MeVc$. Blue dashed: $p$--$\LHe$ without the $\Lambda N$
interaction, which almost coincides with $p$--$\alpha$; the peak shift is therefore due to
the $\Lambda N$ interaction. Green dotted: the factorised estimate $C_{p\alpha}C_{p\Lambda}$,
which reproduces the scale of the enhancement but leaves the peak near $54\MeVc$. Orange:
the full calculation at the free NLO19(600) strength, $s=1$. Blue solid and red dashed: the
calibrated strength, charge symmetric and CSB1. Every $p$--$\LHe$ curve with the $\Lambda N$
interaction resolves the $1^{-}$--$2^{-}$ doublet into two maxima; Sec.~\ref{sec:exp-res}
shows what a measurement would see instead. The vertical scales differ: from left to right the peak height falls by almost
an order of magnitude, while its position moves by less than $0.1\MeVc$.}
\label{fig:corr}
\end{figure*}

Figure~\ref{fig:corr} shows $C(\kstar)$ for Gaussian source radii $R=1.3$, $2.5$ and
$4.0$~fm, which span the range from high-multiplicity $pp$ to peripheral A--A collisions.
All three have the same structure: strong Coulomb suppression below $\kstar\approx15\MeVc$,
a pronounced resonant peak near $30\MeVc$, and a return to unity above
$\approx100\MeVc$.

The peak position is the robust observable. Without the $\Lambda N$ interaction, the peak lies
at $\simeq49\MeVc$, close to the $p$--$\alpha$ peak at $51.6\MeVc$. The folded $\Lambda N$
interaction moves it to about $31\MeVc$ at the calibrated strength, and to about $39\MeVc$ at
$s=0.6$. The position shifts by less than
$0.1\MeVc$ between $R=1.3$ and $4.0$~fm, whereas the height changes by an order of
magnitude. We define the predicted peak position as
\begin{equation}
  \kstar_{\rm peak}=\sqrt{2\mu E_{R}},
  \label{eq:kpeak}
\end{equation}
with $\mu=785.9$~MeV. Numerically, the maximum of the resonant $1^{-}$ channel agrees with
$\kstar_{\rm peak}$ to within $1\%$ for strengths in the calibrated band. The relation
improves as the state narrows, to $0.2\%$ at $s=1.2$, where the state lies only $0.26$~MeV
above threshold. It degrades as the state broadens, to $1.5\%$ at $s=0.6$ and $3.0\%$
without the $\Lambda N$ interaction; Sec.~\ref{sec:piN} explains why. Inverting
Eq.~(\ref{eq:kpeak}) gives $dE_{R}/d\kstar=39$~keV per $\MeVc$ at the calibrated point,
which sets the accuracy target for an experiment (Sec.~\ref{sec:exp}).

The maximum of the total $C(\kstar)$ lies above that of the $1^{-}$ channel, even before any
experimental effect is included. At the calibrated CSB1 strength and $R=2.5$~fm, the total
peaks at $31.25\MeVc$, $0.50\MeVc$ above $\sqrt{2\mu E_{R}}=30.76\MeVc$. Of this
displacement, $0.23\MeVc$ is the shift of the $1^{-}$ channel itself (Sec.~\ref{sec:piN}),
and $0.26\MeVc$ is the pull of the unresolved $2^{-}$ state: without the $2^{-}$ channel,
the maximum is at $30.99\MeVc$. In the charge-symmetric case the $2^{-}$ lies further away,
and the total displacement is only $0.28\MeVc$.

Finite momentum resolution displaces the maximum much further. This experimental effect is
treated in Sec.~\ref{sec:exp-res}; at $\sigma_{\kstar}=3\MeVc$, for example, the maximum
moves to $33.14\MeVc$, $2.4\MeVc$ above $\kstar_{\rm peak}$.

Of the three properties of the peak --- position, height and integrated excess --- the height
is the least robust. At the calibrated CSB1 strength it is $44.6$, $15.6$ and $6.0$ at
$R=1.3$, $2.5$ and $4.0$~fm ($50.5$, $17.4$ and $6.6$ in the charge-symmetric case, which has
the narrower $1^{-}$). The height also suffers most from finite resolution
(Sec.~\ref{sec:exp-res}). Peak \emph{position} and integrated \emph{excess} are therefore the
quantities to quote.

Figure~\ref{fig:corr} also shows the factorised estimate
$C_{p\alpha}(\kstar\mu_{p\alpha}/\mu)\,C_{p\Lambda}(\kstar\mu_{p\Lambda}/\mu)$, which assumes
that the proton scatters independently off the two constituents. This estimate reproduces the
scale of the enhancement but leaves the peak near $54\MeVc$. A product of independent
two-body wave functions cannot move a pole; moving the pole requires the $p\alpha$ and
$p\Lambda$ interactions to act coherently. The estimate is useful only as a null hypothesis that
defines the coherent effect of the $\Lambda$.

\subsection{Partial-wave decomposition}
\label{sec:pw}

\begin{figure*}[tb]
\centering
\includegraphics[width=\textwidth]{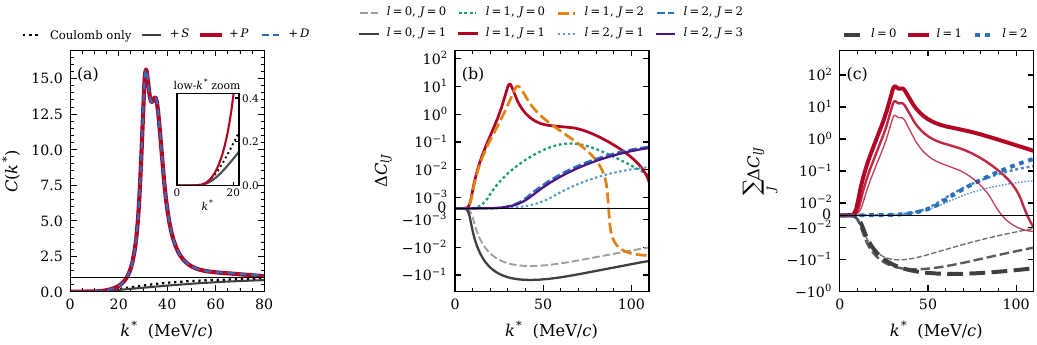}
\caption{Exact partial-wave decomposition of the $p$--$\LHe$ correlation function,
Eq.~(\ref{eq:decomp}), at the calibrated CSB1 strength and without momentum resolution.
(a) Cumulative build-up at $R=2.5$~fm. The $P$ wave creates the entire peak; the $D$ wave is
invisible on this scale. The inset shows the low-$\kstar$ region, where the $S$ wave is the
only strong contribution and pushes the curve \emph{below} the Coulomb baseline.
(b) Individual channel contributions $\Delta C_{lJ}$ at $R=2.5$~fm on a symmetric-log scale,
which shows the sign change of each channel above the peak. The $1^{-}$ and $2^{-}$ channels
peak $4.6\MeVc$ apart.
(c) Contributions summed over $J$ within each partial wave, for the three source radii; line
thickness decreases with increasing $R$. Between $R=1.3$ and $4.0$~fm the $P$-wave term falls
by a factor of eight, while the $S$-wave term is almost unchanged.}
\label{fig:pw}
\end{figure*}

Figure~\ref{fig:pw} and Table~\ref{tab:pw} present the central result of this work: the
exact decomposition of $C(\kstar)$ according to Eq.~(\ref{eq:decomp}) at the calibrated
strength.

At $R=2.5$~fm the $P$ wave carries $98.1\%$ of the peak value of $C$, and the $\LLi$ $1^{-}$
channel alone $76.6\%$. Relative to the enhancement over the Coulomb baseline, the two
fractions are $101.1\%$ and $79.0\%$; the $P$ wave exceeds the whole enhancement because the
$S$ wave cancels part of it. The $P$-wave share of $C$ is $99.5\%$ at $R=1.3$~fm and $91.6\%$
at $R=4.0$~fm. Only the $1^{-}$ share depends appreciably on the $\Lambda N$ scenario: at
$R=1.3$~fm the $1^{-}$ carries $79.3\%$ of the $P$ wave with CSB1 and $88.9\%$ with charge
symmetry, where the $2^{-}$ lies further away. The $2^{-}$ channel peaks at higher momentum.
At the calibrated CSB1 strength it peaks at $35.63\MeVc$ and the $1^{-}$ at $30.99\MeVc$, a
separation of $4.64\MeVc$. In the charge-symmetric case the two maxima are at $35.56$ and
$29.68\MeVc$, a separation of $5.88\MeVc$. The overlap of the two channels produces the double-humped shape in
Figs.~\ref{fig:corr} and~\ref{fig:pw}. The $0^{-}$ channel, which is pure $P_{1/2}$,
contributes only a broad bump of height $0.089$ at $65\MeVc$. The corresponding resonance
lies at $2.41$~MeV with $\Gamma\simeq5$~MeV, i.e.\ at $\sqrt{2\mu E_{R}}=61.5\MeVc$; the bump
lies $3.8\MeVc$ higher, the broad-resonance limit of the shift discussed in
Sec.~\ref{sec:piN}.

The $1^{-}$ lies below the $2^{-}$, by $187$~keV at the calibrated CSB1 strength and by
$232$~keV in the charge-symmetric case. This ordering is an \emph{input assumption}, not a
result. The $1^{-}$ is two thirds singlet and one third triplet, while the $2^{-}$ is pure
triplet, so the splitting follows from the singlet--triplet difference of the $\Lambda N$
interaction, $a_{s}=-2.91$ against $a_{t}=-1.41$~fm in NLO19(600). Nothing in our fit
constrains this difference: the $\LHesix$ binding energy fixes a spin-averaged strength, and
the $\Lambda N$ scattering lengths are taken from the chiral interaction. A spin dependence
of the opposite sign would invert the doublet without affecting any observable used in the
calibration. Whether a correlation measurement could determine the ordering depends on the
momentum resolution (Sec.~\ref{sec:exp-res}).

The $S$ wave contributes a small \emph{negative} term, $\Delta C_{0^{+}}+\Delta
C_{1^{+}}=-0.167$ at the peak, about $1\%$ of the peak height. This is Pauli repulsion: the
Pauli-forbidden $0s$ state forces a node in the interior $S$-wave function, so the $S$-wave
phase shift is effectively repulsive and pushes $C(\kstar)$ \emph{below} the pure-Coulomb
baseline, as the cumulative panel of Fig.~\ref{fig:pw} shows. The $D$ and $F$ waves are
negligible at the peak (Sec.~\ref{sec:numerics}); Sec.~\ref{sec:dip} discusses higher
momenta, where they are not.

Between $R=1.3$ and
$4.0$~fm, the $P$-wave term falls from $44.4$ to $5.5$, while the $S$-wave term stays
between $-0.10$ and $-0.17$. A smaller source probes shorter distances and therefore
amplifies the narrow resonance much more than the smooth $S$-wave background. This is why
small collision systems favour this measurement.

\begin{table*}[tb]
\caption{Partial-wave decomposition of $C(\kstar)$ at the peak, $\kstar=31.25\MeVc$, at the
calibrated CSB1 $\Lambda N$ strength $s=0.891$. Entries are the additive contributions
$\Delta C_{lJ}$ of Eq.~(\ref{eq:dC}); the last column gives the maximum of each channel over
$\kstar$, and the momentum at which it occurs, for $R=2.5$~fm. The sum of all entries
reproduces the directly computed $C(\kstar)$ to machine precision.}
\label{tab:pw}
\begin{ruledtabular}
\small
\begin{tabular}{lrrrl}
Channel & $R=1.3$~fm & $R=2.5$~fm & $R=4.0$~fm & channel maximum ($R=2.5$~fm)\\
\colrule
Coulomb baseline      & $0.358$  & $0.469$  & $0.603$ & ---\\
$S$-wave $0^{+}$ (singlet) & $-0.036$ & $-0.040$ & $-0.024$ & $-0.048$ at $43\MeVc$\\
$S$-wave $1^{+}$ (triplet) & $-0.115$ & $-0.127$ & $-0.076$ & $-0.152$ at $43\MeVc$\\
$P$-wave $0^{-}$      & $+0.013$ & $+0.009$ & $+0.006$ & $+0.089$ at $65\MeVc$\\
$P$-wave $1^{-}$      & $\bm{+35.22}$ & $\bm{+11.98}$ & $\bm{+4.21}$ & $+12.14$ at $31.0\MeVc$\\
$P$-wave $2^{-}$      & $+9.20$  & $+3.35$  & $+1.31$ & $+10.23$ at $35.6\MeVc$\\
$D$ waves (sum)       & $+0.0005$ & $+0.0005$ & $+0.0004$ & $+0.17$ at $128\MeVc$\\
\colrule
Total $C(\kstar)$     & $44.64$  & $15.64$  & $6.03$  & \\
\end{tabular}
\end{ruledtabular}
\end{table*}

\subsection{Peak position and resonance width}
\label{sec:piN}

Higher partial waves also shape other correlation functions: for the deuteron--deuteron
system the $P$ waves are indispensable~\cite{Ge2025}. In the present case, however, the $P$
wave carries the entire positive enhancement at all momenta, because the $S$-wave
contribution is negative everywhere (Sec.~\ref{sec:pw}).

How well the channel maximum agrees with $\kstar_{\rm peak}$ depends on the width of the
resonance. Across a resonance the phase shift passes through $\pi/2$ over an energy
interval of order $\Gamma$. Over that interval the Coulomb and centrifugal factors also
vary, so the maximum of a resonant channel is displaced from $\sqrt{2\mu E_{R}}$ by an
amount that grows with the width. The $\LLi$ $1^{-}$ and the $^{5}$Li $3/2^{-}$ of the model show
this directly. At
the calibrated CSB1 strength, $\Delta C_{1^{-}}$, computed alone on a fine momentum grid,
peaks at $30.99\MeVc$, $0.23\MeVc$ or $0.8\%$ above $\sqrt{2\mu E_{R}}=30.76\MeVc$. For the
broad $^{5}$Li $3/2^{-}$ state ($\Gamma=1.33$~MeV), the $p$--$\alpha$ channel peaks at
$51.29\MeVc$, $3.3\%$ above the value $49.67\MeVc$ computed from the $S$-matrix pole. A width
$7.5$ times larger thus produces a relative displacement four times larger. The peak
position is a good measure of $E_{R}$ for the narrow $\LLi$ $1^{-}$, but for a broad
resonance the displacement must be calculated.

\subsection{The suppression above the peak}
\label{sec:dip}

Liu \textit{et al.}~\cite{Liu2024} predict a minimum \emph{above} the peak of a resonant
correlation function. Writing the exterior radial function as $u_{l}=\cos\delta\,F_{l}+\sin\delta\,G_{l}$
gives
\begin{equation}
  |u_{l}|^{2}-F_{l}^{2}=\sin^{2}\!\delta\,(G_{l}^{2}-F_{l}^{2})
  +\sin2\delta\;F_{l}G_{l},
  \label{eq:dipdecomp}
\end{equation}
and the second term changes sign as $\delta$ passes $\pi/2$; they identified this mechanism
for the $Z_{c}(3900)$ and $Z_{cs}(3985)$. Our calculation contains the same structure, but
its consequences differ for the individual channels and for the total $C(\kstar)$.

\begin{figure}[tb]
\centering
\includegraphics[width=\columnwidth]{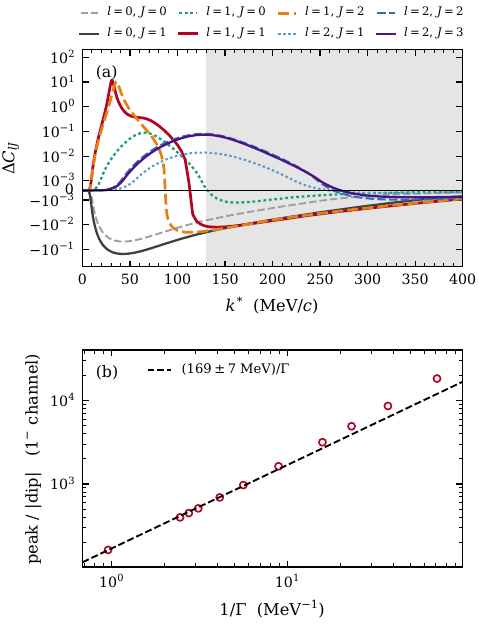}
\caption{The suppression above the peak, at $R=2.5$~fm and without momentum resolution.
(a) Channel contributions $\Delta C_{lJ}$ on a symmetric-log scale at the calibrated CSB1
strength. Both $P$-wave channels change sign above the peak, the $1^{-}$ at $113\MeVc$ and
the $2^{-}$ at $87\MeVc$; these crossings are exact. The total $C(\kstar)$ is not drawn, because
the partial-wave sum is not converged above $100\MeVc$ ($l=4$ is $71\%$ of $l=3$ at
$240\MeVc$) and the momentum is far above the $p+\alpha+\Lambda$ breakup threshold. In the
shaded region, above $\approx130\MeVc$, the $D$ wave exceeds $C-1$.
(b) Ratio of peak height to dip depth in the $1^{-}$ channel against $1/\Gamma$, on
logarithmic axes, from a scan of the $\Lambda N$ strength covering a factor of $75$ in
$\Gamma$; peak heights are taken on a fine momentum grid. The dashed line is
Eq.~(\ref{eq:dipratio}), fitted to the seven points with $\Gamma\ge0.11$~MeV. The four
narrower states lie increasingly above it, because the peak grows faster than $1/\Gamma$
near threshold.}
\label{fig:dip}
\end{figure}

In the individual channels the structure is present and exactly calculable
(Fig.~\ref{fig:dip}). At $R=2.5$~fm and the calibrated strength, $\Delta C_{1^{-}}$ changes
sign at $113\MeVc$ and reaches $-0.013$ near $140\MeVc$; $\Delta C_{2^{-}}$ changes sign at
$87\MeVc$ and reaches $-0.020$ near $113\MeVc$. These are the quantities that the mechanism
of Eq.~(\ref{eq:dipdecomp}) controls, and they confirm that nothing in our treatment
suppresses the sign change.

For the total $C(\kstar)$ we do not quote a dip depth, for two reasons. First, the
partial-wave sum is not converged in that region. The nuclear interaction is retained for
$l\le3$, which is ample at the peak, but the higher waves grow with momentum: the $D$ wave
reaches $+0.112$ at $100\MeVc$ and $+0.168$ at $130\MeVc$, where it exceeds $C-1$, and the
$F$ wave reaches $+0.027$ at $240\MeVc$, where the next term is still $71\%$ of it. With
$l\le3$, the total does not fall below unity until $264\MeVc$, which shows how little such a
number means. Second, breakup into $p+\alpha+\Lambda$ opens at $\kstar\approx70\MeVc$
(Sec.~\ref{sec:limits}); above that momentum the elastic two-body wave function does not
describe the correlation function. A total-$C$ dip above $100\MeVc$ would therefore be an
artefact of a truncated sum, in a region where the model does not apply.

We scan the $\Lambda N$ strength from $s=0$ to
$s=1.25$, which reduces $\Gamma$ by a factor of $75$, from $1.048$ to $0.014$~MeV. Over this
range the channel dip deepens only by a factor of $1.6$, from $-0.0090$ to $-0.0143$, while
the peak grows by a factor of $180$. For the broader states the two combine into a simple
law,
\begin{equation}
  \frac{\text{peak}}{|\text{dip}|}\simeq\frac{169\pm7\;\mathrm{MeV}}{\Gamma},
  \label{eq:dipratio}
\end{equation}
which holds to within $8\%$ for the seven cases with $\Gamma\ge0.11$~MeV, a factor of nine
in $\Gamma$ that includes the calibrated band. For narrower states the peak grows even
faster: $\Gamma\cdot\text{peak}/|\text{dip}|$ rises to $230$~MeV at $\Gamma=0.027$~MeV and
to $260$~MeV at $\Gamma=0.014$~MeV, where the state lies only $0.21$~MeV above threshold. In
both regimes the growth is in the peak, not in the dip. The Coulomb and centrifugal barrier
does not suppress the dip; it \emph{enhances the peak}, through the factor $1/P_{l}(\kstar)$,
where $P_{l}$ is the barrier penetrability. The confinement that makes the state narrow also makes the resonant
peak tall.

This is the essential difference from the $Z_{c}$ case~\cite{Liu2024}. There the
resonances lie at relative momenta of several hundred $\MeVc$, where $F_{l}$ and $G_{l}$ are
comparable and no Coulomb barrier amplifies the peak, so peak and dip have similar
magnitudes and both are predicted to be observable. In a low-momentum, charged, $P$-wave
system such as $p$--$\LHe$, the dip is $10^{-3}$ of the peak, and the peak position and
integrated excess are the quantities to measure.

\subsection{$S$-wave scattering parameters and the limits of an LL analysis}
\label{sec:ere}

\begin{figure}[tb]
\centering
\includegraphics[width=\columnwidth]{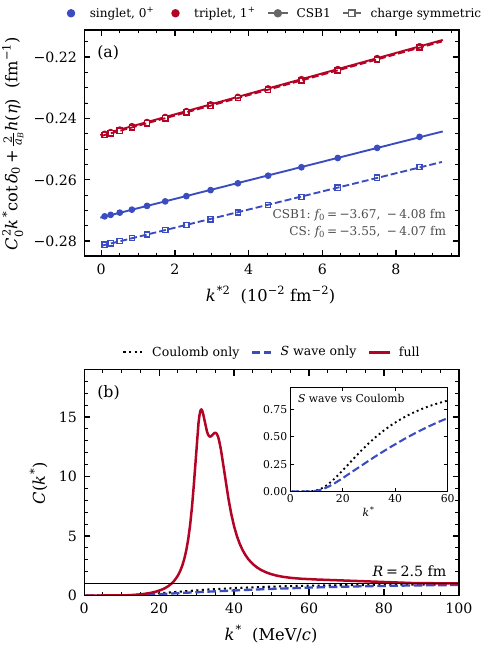}
\caption{(a) The Coulomb-modified effective-range function of Eq.~(\ref{eq:ere}) for the
singlet (blue) and triplet (red) $S$-wave channels, at the calibrated CSB1 strength (filled
circles, solid fits) and with charge symmetry (open squares, dashed fits). Symbols are the
computed phase shifts; lines are two-parameter fits with intercept $1/f_{0}$ and slope
$d_{0}/2$.
(b) The correlation function with only the $S$-wave interaction retained (dashed) never
rises above unity and lies \emph{below} the pure-Coulomb baseline (dotted), whereas the full
result (solid) peaks near $31\MeVc$. The inset shows the two suppressed curves on their own
scale.}
\label{fig:ere}
\end{figure}

Femtoscopic analyses conventionally quote the scattering length $f_{0}$ and effective range
$d_{0}$ of the LL amplitude
$f(\kstar)=\bigl(1/f_{0}+\tfrac12 d_{0}\kstarsq-i\kstar\bigr)^{-1}$, so that $f_{0}=-a$,
where $a$ is the standard scattering length. We extract them from the Coulomb-modified
effective-range expansion
\begin{equation}
  C_{0}^{2}(\eta)\,\kstar\cot\delta_{0}+\frac{2}{a_{B}}h(\eta)
  =\frac{1}{f_{0}}+\frac{1}{2}d_{0}\kstarsq,
  \label{eq:ere}
\end{equation}
with $C_{0}^{2}(\eta)=2\pi\eta/(e^{2\pi\eta}-1)$, $\eta=1/(\kstar a_{B})$ and
$h(\eta)=\mathrm{Re}\,\psi(1+i\eta)-\ln\eta$, fitted over $\kstar=6$--$60\MeVc$.
Figure~\ref{fig:ere}(a) shows that the expansion is linear over this range. As a validation,
the same fit applied to the $p\Lambda$ subsystem recovers its input scattering length to
$0.1\%$ and its effective range to $2\%$.

Table~\ref{tab:ere} collects the results. All values of $f_{0}$ are negative and there is no
$S$-wave bound state: the channel is Pauli-blocked and effectively repulsive, as for
$p\alpha$. Adding the $\Lambda$ reduces $|f_{0}|$ from $5.53$ to $3.67$--$4.08$~fm
($3.55$--$4.07$~fm with charge symmetry): the $\Lambda N$ attraction cancels part of the
Pauli repulsion without changing its sign. The singlet is less repulsive than the triplet,
as expected from the more attractive singlet $\Lambda N$ force.

These parameters, however, \emph{do not describe the measurable signal}.
Figure~\ref{fig:ere}(b) compares the full $C(\kstar)$ with the result obtained when only the
$S$-wave interaction is retained. The $S$-wave-only correlation function never rises above
unity. It lies \emph{below} the pure-Coulomb baseline, and its ratio to that baseline falls
to $0.57$ at $R=1.3$~fm, $0.63$ at $R=2.5$~fm (the case shown) and $0.81$ at $R=4$~fm. An LL
analysis restricted to $S$ waves can at best reproduce this curve; it cannot generate a peak.
Fitting measured $p$--$\LHe$ data with such a parametrisation would return meaningless
parameters. For this system the quantities to extract are the resonance position, obtained
from the peak position through Eq.~(\ref{eq:kpeak}), and the width.

\begin{table*}[tb]
\caption{Coulomb-modified $S$-wave parameters in the femtoscopy (Lednick\'y--Lyuboshitz)
convention, $f_{0}=-a$ and $d_{0}=r_{\rm eff}$, from a two-parameter fit of
Eq.~(\ref{eq:ere}) over $\kstar=6$--$60\MeVc$. All values in fm. Against changes of the fit
range and of the number of shape parameters, $f_{0}$ is stable to better than $0.1\%$ and
$d_{0}$ to about $5\%$. The first two systems have no spin-dependent interaction, so their
two channels are identical; in the remaining rows the spin dependence comes entirely from
the folded $\Lambda N$ force. Boldface marks the calibrated CSB1 values, the default of this
paper (Sec.~\ref{sec:A6}), quoted in Sec.~\ref{sec:summary}. The charge-symmetric row above
it differs appreciably only in the singlet, where the $\Lambda p$ scattering length changes.
The systematics of Sec.~\ref{sec:sys} are evaluated at the CSB1 point.}
\label{tab:ere}
\begin{ruledtabular}
\small
\begin{tabular}{lrrrr}
 & \multicolumn{2}{c}{singlet ($S=0$)} & \multicolumn{2}{c}{triplet ($S=1$)}\\
\cline{2-3}\cline{4-5}
System & $f_{0}$ & $d_{0}$ & $f_{0}$ & $d_{0}$\\
\colrule
$p$--$\alpha$                        & $-5.31$ & $+0.95$ & $-5.31$ & $+0.95$\\
$p$--$\LHe$, no $\Lambda N$          & $-5.53$ & $+0.93$ & $-5.53$ & $+0.93$\\
\colrule
$p$--$\LHe$, free NLO19(600) ($s=1$)      & $-3.37$ & $+0.58$ & $-3.94$ & $+0.64$\\
$p$--$\LHe$, calibrated CS ($s=0.915$) & $-3.55$ & $+0.59$ & $-4.07$ & $+0.67$\\
$p$--$\LHe$, calibrated CSB1 ($s=0.891$) & $\bm{-3.67}$ & $\bm{+0.60}$ & $\bm{-4.08}$ & $\bm{+0.67}$\\
\colrule
$n$--$\LHe$, calibrated CS           & $-1.83$ & $+0.66$ & $-2.01$ & $+0.69$\\
\end{tabular}
\end{ruledtabular}
\end{table*}

\subsection{Charge-symmetry breaking}
\label{sec:csb}

The only empirical information on the $\Lambda n$--$\Lambda p$ difference comes from the
$A=4$ mirror hypernuclei~\cite{Yamamoto2015,Schulz2016,Gazda2016}. We adopt the chiral EFT
``CSB1'' scenario of Haidenbauer, Mei\ss ner and Nogga~\cite{HMN2021} for NLO19(600). It is
fitted to $\Delta E(0^{+})=233\pm92$~keV and $\Delta E(1^{+})=-83\pm94$~keV and yields
$a_{s}^{\Lambda p}=-2.63$~fm, $a_{s}^{\Lambda n}=-3.23$~fm, $a_{t}^{\Lambda p}=-1.47$~fm and
$a_{t}^{\Lambda n}=-1.36$~fm. We keep the Gaussian ranges, refit the depths, and use the
$\Lambda n$ interaction for $\LHesix$ and the $\Lambda p$ interaction for $\LLi$. $\LHe$
itself is unaffected to first order, because its $\alpha$ core has isospin zero.

Refitting the strength to $B_\Lambda(\LHesix)$ gives $s=0.891^{+0.072}_{-0.076}$ and raises
the $\LLi$ $1^{-}$ by $49$~keV, to $E_{R}=0.602$~MeV with $\Gamma=0.178$~MeV; the
pure-triplet $2^{-}$ is essentially unchanged. The mirror splitting thus grows, but only by
$50$--$65$~keV in $A=6$, about a fifth of the $A=4$ value. The valence nucleon occupies a
$P$ orbit with reduced overlap with the $\Lambda$, and only one nucleon carries the charge
difference. This dilution is consistent with the small CSB found in the $A=7$ systems in
the NCSM~\cite{Le2023}.

In the correlation function, CSB shifts the maximum of the $1^{-}$ channel by $+1.3\MeVc$. The
maximum at $\sigma_{\kstar}=3\MeVc$ moves by the same amount, and the unsmeared total maximum
by $+1.5\MeVc$; the smeared peak height is essentially unchanged. The shift comes from the
predominantly singlet $1^{-}$ state. The factorised estimate $C_{p\alpha}C_{p\Lambda}$ does
not register it at all: in the spin-averaged $p\Lambda$ correlation, the weakening of the
$\Lambda p$ singlet (weight $1/4$) and the slight strengthening of the triplet (weight $3/4$)
relative to charge symmetry nearly cancel.

A $p$--$\LHe$ measurement nevertheless cannot be used to study CSB, and not merely because
the effect is small. \emph{The CSB shift is degenerate with the calibration.} From
Table~\ref{tab:A6}, $dE_{R}/ds\simeq-1.09$~MeV per unit strength, so a shift of $+49$~keV in
$E_{R}$ is equivalent to a change $\Delta s\simeq0.045$ in the overall $\Lambda N$ strength.
This is well inside the range $^{+0.075}_{-0.078}$ allowed by the $\pm0.10$~MeV uncertainty
of $B_\Lambda(\LHesix)$. CSB therefore changes the prediction in exactly the direction that
$s$ absorbs, which is why the charge-symmetric and CSB1 bands in Fig.~\ref{fig:mirror}
overlap. However precise, a measurement of the peak position cannot separate the two unless
the overall strength is fixed independently to better than $5\%$.

A second, weaker point concerns the absolute accuracy of the pole. Because $\LHesix$ is
bound and $\LLi$ is a resonance, part of the model dependence does not cancel between them:
the $\LLi$ pole also depends on the $^{5}$Li analog energy and on the barrier
penetrability, which the bound state does not constrain. The $N\alpha$ range scan of
Sec.~\ref{sec:nalpharange} shows this directly. There $s$ is \emph{refitted} to
$B_\Lambda(\LHesix)$ for each choice of ranges and varies over $0.771$--$1.066$, yet the
$\LLi$ $1^{-}$ position still moves from $0.564$ to $0.650$~MeV, i.e.\ by $\pm43$~keV. The
reason is that the $^{5}$Li analog energy drifts from $1.58$ to $1.72$~MeV, while
$B_\Lambda(\LHesix)$ is fixed by construction. This residual uncertainty concerns the
absolute position. The CSB difference is evaluated at fixed $N\alpha$ input and is more stable
than $\pm43$~keV; we quote this scale because any claimed $50$~keV effect would have to be
defended against it.

We therefore treat the difference between the CS and CSB1 results as a variation of the
prediction, not as an observable. Establishing CSB in $A=6$ from the continuum would require fixing the overall
$\Lambda N$ strength independently of the mirror binding energy, which is a different
measurement from the one proposed here.

\section{Systematic uncertainties}
\label{sec:sys}

Three inputs of the model are not determined by the calibration observables: the two fixed
$N\alpha$ ranges, the $\Lambda N$ form factor, and the choice among chiral $\Lambda N$
solutions. For each variation we repeat the entire chain --- refit of the two-body depths,
recalibration of $s$ on $B_\Lambda(\LHesix)$, and re-extraction of the $\LLi$ resonance ---
and propagate the result to the predicted peak position $\kstar_{\rm peak}=\sqrt{2\mu E_{R}}$.

\subsection{$N\alpha$ ranges}
\label{sec:nalpharange}
\label{sec:sys-nalpha}

Table~\ref{tab:sys-nalpha} varies $b_{C}$ and $b_{LS}$ by $\pm0.2$~fm about their adopted
values; each time, $V_{0}$ and $V_{LS}$ are refitted to the same two neutron observables.
The spin--orbit range is almost irrelevant: $\pm0.2$~fm moves the peak by $\mp0.12\MeVc$.
The central range matters much more and moves the peak over $29.8$--$32.0\MeVc$. The reason
is a strong parameter correlation. A shorter central range requires a deeper $V_{0}$ to
reproduce $a_{S}(n\alpha)$, and the deeper $V_{0}$ already supplies much of the $P$-wave
attraction; the spin--orbit strength that places $^{5}$He at $0.80$~MeV therefore varies
from $8.2$ to $38.5$~MeV across the range considered. The $^{5}$Li pole follows this
variation only weakly ($E_{R}=1.58$--$1.72$~MeV, bracketing the data-derived pole at
$1.69$~MeV~\cite{Tilley2002}), so it does not discriminate sharply between these choices.
We therefore assign half the spread, $\pm1.1\MeVc$, as a systematic uncertainty. Of all
budget entries, this is the largest that a better $N\alpha$ potential could remove
(Sec.~\ref{sec:limits}).

\subsection{$\Lambda N$ form factor}
\label{sec:sys-lamN}

Table~\ref{tab:sys-lamN} scales the $\Lambda N$ Gaussian ranges by $\pm25\%$ at
\emph{fixed} scattering lengths, with the depths refitted. The singlet effective range then
changes from $2.16$ to $3.82$~fm, nearly a factor of two and far outside the chiral
uncertainty, and with it the off-shell behaviour that controls how the folded potential acts
in the $p$--$\LHe$ $P$ waves. The calibrated strength compensates almost exactly: it moves
from $0.963$ to $0.864$, while the resonance position shifts by $2$~keV and the peak position
by $0.04\MeVc$. Once the spin-weighted strength fixed by $B_\Lambda(\LHesix)$ is calibrated,
the $\LLi$ prediction is thus essentially insensitive to the $\Lambda N$ form factor. This answers the concern
raised in Sec.~\ref{sec:input}: an $S$-wave-constrained $\Lambda N$ input suffices here not
because $\Lambda N$ $P$ waves are negligible in general, but because folding followed by
calibration leaves the $p$--$\LHe$ potential with essentially one effective degree of
freedom.

\subsection{Choice of chiral $\Lambda N$ solution}
\label{sec:sys-chiral}

Table~\ref{tab:sys-chiral} repeats the calculation for the NLO13~\cite{NLO13} and
NLO19~\cite{HMN2020} chiral $YN$ interactions at cutoffs of $500$--$650$~MeV, each with the
$\Lambda p$ and $\Lambda n$ scattering lengths of the CSB1 scenario~\cite{HMN2021}. Without
recalibration ($s=1$), all eight solutions overbind $\LHesix$, giving
$B_\Lambda=4.33$--$4.36$~MeV against the measured $4.18\pm0.10$~MeV, and place the $\LLi$
$1^{-}$ state at $0.476$--$0.489$~MeV, i.e.\ $\kstar_{\rm peak}=27.4$--$27.7\MeVc$. After
recalibration the strengths lie in a narrow band, $s=0.877$--$0.891$, and the predictions
converge further: $E_{R}$ spans only $15$~keV and $\kstar_{\rm peak}$ only $0.4\MeVc$. The
spread between NLO13 and NLO19 is comparable to the cutoff spread within either family.

This convergence has a structural reason. The $\LHesix$ $1^{-}$ ground state, on which $s$
is fitted, and the predicted $\LLi$ $1^{-}$ state are mirror states with the same spin
content: two thirds singlet and one third triplet in the $j_{p}=3/2$ channel. The calibration
therefore fixes exactly the combination $\tfrac23V_{s}+\tfrac13V_{t}$ on which the prediction
depends. Any change that a chiral solution makes to this combination is absorbed; only the
orthogonal part remains, and it accounts for the $15$~keV spread. The same argument explains
why the $\Lambda N$ form factor has almost no effect (Table~\ref{tab:sys-lamN}), and why the
pure-triplet $2^{-}$ state is \emph{not} protected in this way (Sec.~\ref{sec:budget}).

\subsection{Uncertainty budget}
\label{sec:budget}

Combining these variations with the experimental uncertainty of $B_\Lambda(\LHesix)$
(Sec.~\ref{sec:A6}), the uncertainty budget of the predicted peak position is
\begin{align}
  \text{$B_\Lambda(\LHesix)$, }\pm0.10\text{ MeV} &: \;\pm2.0\MeVc,\nonumber\\
  \text{$N\alpha$ central range} &: \;\pm1.1\MeVc,\nonumber\\
  \text{chiral $\Lambda N$ solution} &: \;\pm0.2\MeVc,\nonumber\\
  \text{$\Lambda N$ form factor} &: \;\pm0.02\MeVc,\nonumber\\
  \text{$N\alpha$ spin--orbit range} &: \;\pm0.12\MeVc,\nonumber\\
  \text{$^{5}$He energy in $B_{n}$} &: \;-1.3\MeVc\ \text{(one-sided)}.\nonumber
\end{align}
The last entry concerns the $^{5}$He energy used to convert $B_\Lambda(\LHesix)$ into the
neutron separation energy $B_{n}$ on which $s$ is fitted (Sec.~\ref{sec:A6}). We adopt the
evaluated pole position, $0.80$~MeV~\cite{Tilley2002}; the mass-table value,
$0.735$~MeV~\cite{AME2020}, would move $s$ from $0.891$ to $0.939$ and $\kstar_{\rm peak}$
from $30.8$ to $29.5\MeVc$. This shift is one-sided; we carry it at its full size and, for
simplicity, add it in quadrature on both sides. Refitting the $N\alpha$ potential to the speed-plot value
of the data, $0.77$~MeV, with $B_{n}$ fixed moves $\kstar_{\rm peak}$ by only $0.04\MeVc$, so
the choice of fit target itself is negligible. The experimental $\LHesix$ binding energy is
the dominant source of uncertainty, followed by the $^{5}$He energy in $B_{n}$ and the
$N\alpha$ central range. The $\Lambda N$ sector contributes least, because it is the sector
that the calibration constrains. Added in quadrature, the entries give
$\kstar_{\rm peak}=30.8\pm2.6\MeVc$, or $E_{R}(\LLi\,1^{-})=0.60^{+0.11}_{-0.10}$~MeV. A
more precise prediction therefore requires a better $\LHesix$ binding energy rather than a
better $\Lambda N$ interaction.

Three further uncertainties are either one-sided or not expressible as a shift of the peak,
and we list them separately. First, the frozen core is expected to underbind relative to a
full three-body treatment, so $s=0.891$ is an upper estimate and the predicted $E_{R}$ a
lower one. The $45$~keV deficit of the model $^{5}$Li pole (Sec.~\ref{sec:A6}) points the
same way, by about $45$~keV. Second, the width has a much larger relative uncertainty
than the position: the $\pm0.10$~MeV on $B_\Lambda(\LHesix)$ alone gives
$\Gamma(\LLi\,1^{-})=0.178^{+0.053}_{-0.045}$~MeV, a $+30/{-25}\%$ band, which we quote. The
$p\alpha$ sector does not require a wider band: the model's $P_{3/2}$ pole widths exceed the
data-derived ones by $8\%$ for both $^{5}$He and $^{5}$Li (Sec.~\ref{sec:input}), which
suggests that $\Gamma(\LLi)$ may be high by a similar amount, well inside the band. Third,
the $2^{-}$ state is pure triplet and is therefore \emph{not} constrained by the calibration
(Sec.~\ref{sec:sys-chiral}). We
moved the $2^{-}$ to Hiyama's $0.97$~MeV by weakening the triplet $\Lambda N$ strength and
refitting the singlet so that $B_\Lambda(\LHesix)$ is unchanged. This raises the $1^{-}$ by
only $8$~keV, but moves the merged peak by $-0.8\MeVc$ at $\sigma_{\kstar}=3\MeVc$, because a
more distant $2^{-}$ is broader and pulls the merged peak less. At $\sigma_{\kstar}=5\MeVc$,
where the doublet is fully merged, the merged peak instead follows the $2^{-}$ upward, by
$+0.7\MeVc$. This dependence is not included in the budget, and it is the hardest to reduce within
the model.

\begin{table}[tb]
\caption{Sensitivity to the two fixed $N\alpha$ ranges. In every row the depths are refitted
to $a_{S}(n\alpha)=2.46$~fm and $E_{R}(^{5}\mathrm{He})=0.80$~MeV. Energies and depths in
MeV, ranges in fm, $\kstar_{\rm peak}$ in MeV/$c$. The $^{5}$Li columns give the $S$-matrix
pole, for comparison with the data-derived pole at $1.69$~MeV,
$\Gamma=1.23$~MeV~\cite{CsotoHale1997}. The first row is the adopted parametrisation.}
\label{tab:sys-nalpha}
\begin{ruledtabular}
\small
\begin{tabular}{cccccccc}
$b_{C}$ & $b_{LS}$ & $V_{0}$ & $V_{LS}$ & $^{5}$Li $E_{R}$ & $^{5}$Li $\Gamma$ & $s$ &
$\kstar_{\rm peak}$\\
\colrule
2.30 & 1.70 & $-47.69$ & $-23.47$ & 1.65 & 1.33 & 0.891 & 30.8\\
2.10 & 1.70 & $-53.45$ & $-38.45$ & 1.72 & 1.25 & 0.771 & 32.0\\
2.50 & 1.70 & $-42.55$ & $-8.16$  & 1.58 & 1.42 & 1.066 & 29.8\\
2.30 & 1.50 & $-47.69$ & $-33.26$ & 1.65 & 1.33 & 0.869 & 30.9\\
2.30 & 1.90 & $-47.69$ & $-17.56$ & 1.64 & 1.34 & 0.911 & 30.6\\
\end{tabular}
\end{ruledtabular}
\end{table}

\begin{table}[tb]
\caption{Sensitivity to the $\Lambda N$ form factor. The Gaussian ranges are scaled by the
factor in the first column at fixed scattering lengths, so the singlet effective range
$r_{s}$ changes while $a_{s}=-2.632$~fm is held. Middle row: adopted.}
\label{tab:sys-lamN}
\begin{ruledtabular}
\small
\begin{tabular}{cccccc}
range factor & $b_{s}$ (fm) & $r_{s}$ (fm) & $s$ & $E_{R}$ (MeV) & $\kstar_{\rm peak}$\\
\colrule
0.80 & 1.147 & 2.16 & 0.963 & 0.600 & 30.7\\
1.00 & 1.434 & 2.86 & 0.891 & 0.602 & 30.8\\
1.25 & 1.792 & 3.82 & 0.864 & 0.602 & 30.8\\
\end{tabular}
\end{ruledtabular}
\end{table}

\begin{table*}[tb]
\caption{The full chain repeated for eight chiral $YN$ solutions. $a_{s}$ and $a_{t}$ are the
$\Lambda p$ scattering lengths of the CSB1 scenario~\cite{HMN2021}; the corresponding
$\Lambda n$ values are used for $\LHesix$. Left block: without recalibration, $s=1$. Right
block: after recalibration on $B_\Lambda(\LHesix)=4.18$~MeV. Energies in MeV, lengths in
fm, momenta in MeV/$c$.}
\label{tab:sys-chiral}
\begin{ruledtabular}
\small
\begin{tabular}{lcccccccccc}
 & & & \multicolumn{3}{c}{$s=1$ (uncalibrated)} & & \multicolumn{3}{c}{calibrated}\\
Solution & $a_{s}$ & $a_{t}$ & $B_\Lambda(\LHesix)$ & $E_{R}$ & $\kstar_{\rm peak}$ & $s$ &
$E_{R}$ & $\Gamma$ & $\kstar_{\rm peak}$\\
\colrule
NLO13(500) & $-2.604$ & $-1.647$ & 4.356 & 0.477 & 27.4 & 0.877 & 0.609 & 0.183 & 30.9\\
NLO13(550) & $-2.586$ & $-1.551$ & 4.350 & 0.486 & 27.7 & 0.880 & 0.613 & 0.186 & 31.0\\
NLO13(600) & $-2.588$ & $-1.573$ & 4.352 & 0.484 & 27.6 & 0.879 & 0.613 & 0.185 & 31.0\\
NLO13(650) & $-2.592$ & $-1.538$ & 4.346 & 0.487 & 27.7 & 0.883 & 0.611 & 0.184 & 31.0\\
NLO19(500) & $-2.649$ & $-1.580$ & 4.339 & 0.476 & 27.4 & 0.887 & 0.598 & 0.176 & 30.7\\
NLO19(550) & $-2.640$ & $-1.524$ & 4.333 & 0.482 & 27.5 & 0.890 & 0.599 & 0.176 & 30.7\\
NLO19(600) & $-2.632$ & $-1.473$ & 4.331 & 0.487 & 27.7 & 0.891 & 0.602 & 0.178 & 30.8\\
NLO19(650) & $-2.620$ & $-1.464$ & 4.331 & 0.489 & 27.7 & 0.891 & 0.604 & 0.179 & 30.8\\
\end{tabular}
\end{ruledtabular}
\end{table*}

\section{Relation to previous work}

Table~\ref{tab:compare} places the present calculation among existing treatments of the
$A=6$ hypernuclei. As a structure calculation our model is the least sophisticated of the
group; its purpose is different.

The three-body cluster calculations~\cite{Motoba1983,Hiyama1996,Filikhin2014} solve the
$\alpha+\Lambda+N$ problem essentially exactly, including the rearrangement of the $\Lambda$
and the core. The closest to the present work is that of Hiyama
\textit{et al.}~\cite{Hiyama1996}. It solves the full three-body problem with the Gaussian
expansion method and uses the KKNN $\alpha N$ potential~\cite{KKNN1979}, which our
single-Gaussian $N\alpha$ potential resembles in range and depth. The agreement
with our frozen-core treatment is better than the difference in sophistication would
suggest. For the $\LHesix$ ground state they obtain $-0.25$~MeV relative to $n+\LHe$, against
our $-0.260$~MeV, to which $s$ is fitted; and for the $\LLi$ mirror $+0.58$~MeV, against our $0.55$~MeV ($0.60$~MeV
with CSB). Their $2^{-}$ states lie at $+0.11$ and $+0.97$~MeV, against our $-0.002$ and
$+0.79$~MeV. The doublet splitting is therefore the one quantity in which the two
calculations differ appreciably: $0.39$~MeV ($\LLi$) and $0.36$~MeV ($\LHesix$) in their
calculation, against our $0.23$ and $0.26$~MeV. The halo radii agree to $7$--$9\%$
(Sec.~\ref{sec:A6}); the remaining difference is due to the core contraction, which a frozen
core cannot reproduce. Their bound-state expansion is not designed to give phase shifts, so
it provides neither the widths of the $\LLi$ states nor any continuum observable. The \emph{ab initio}
calculations~\cite{Wirth2014,WirthRoth2016,Le2020,Le2023} go further and treat all six
baryons with realistic chiral interactions and explicit $\Sigma$ degrees of freedom. Our
folding model differs from them physically by neglecting core polarisation and genuine
three-body forces; both are absorbed into the empirical factor $s$ ($0.891$ with CSB1,
$0.915$ with charge symmetry). The microscopic calculations find, from first principles, that repulsive
$\Lambda NN$ or SRG-induced $YNN$ terms reduce the effective attraction by a similar amount
and in the same direction~\cite{WirthRoth2016,Lonardoni2014,Contessi2018}; our factor $s<1$
is the phenomenological counterpart of that result.

What the present work adds is the observable. Of the previous studies, only
Oishi~\cite{Oishi2018} addresses the $\LLi$ continuum directly, through a time-dependent
three-body treatment of proton emission, and that work established the sensitivity of the
emission observables to the $p\Lambda$ scattering length. It does not provide a correlation
function: it contains no emission source, no Coulomb-distorted two-particle wave function
folded over it, and no partial-wave-resolved $C(\kstar)$. To our knowledge, the quantity
measured in a femtoscopy experiment --- $C(\kstar)$ as a function of source size, with the
channel content made explicit and the momentum resolution folded in --- has not been
computed for any hypernucleus--nucleon pair. A simplified but complete treatment is
therefore useful at this stage.

\begin{table*}[tb]
\caption{Theoretical treatments of the $A=6$ hypernuclei. ``Continuum'' indicates whether the
$\LLi$ resonance is treated as such; ``$C(\kstar)$'' whether a femtoscopic correlation
function is computed.}
\label{tab:compare}
\begin{ruledtabular}
\small
\begin{tabular}{llccl}
Work & Method & Continuum & $C(\kstar)$ & Emphasis\\
\colrule
Motoba \textit{et al.}~\cite{Motoba1983} & microscopic $\alpha+\chi+\Lambda$ cluster & no & no & spectra, EM transitions\\
Hiyama \textit{et al.}~\cite{Hiyama1996} & $\alpha+\Lambda+N$, Gaussian expansion & no & no & binding, halo/skin structure\\
Filikhin \textit{et al.}~\cite{Filikhin2014} & configuration-space Faddeev & no & no & doublet splitting, 3-body force\\
Oishi~\cite{Oishi2018} & time-dependent 3-body & yes & no & 1$p$ emission, $a_{p\Lambda}$ sensitivity\\
Wirth \textit{et al.}~\cite{Wirth2014,WirthRoth2016} & NCSM, chiral $YN$, $\Sigma$ explicit & no & no & \emph{ab initio} spectra, induced $YNN$\\
Le \textit{et al.}~\cite{Le2020,Le2023} & Jacobi NCSM, NLO chiral $YN$ & no & no & separation energies, CSB\\
Contessi \textit{et al.}~\cite{Contessi2018} & pionless EFT + SVM & no & no & overbinding, $\Lambda NN$ LECs\\
Lonardoni \textit{et al.}~\cite{Lonardoni2014} & AFDMC & no & no & $\Lambda NN$ from $\Lambda$ separation energies\\
\textbf{This work} & $\alpha+\Lambda$ folding, frozen core & \textbf{yes} & \textbf{yes} & \textbf{partial-wave-resolved femtoscopy}\\
\end{tabular}
\end{ruledtabular}
\end{table*}

\section{How the measurement can be made}
\label{sec:exp}

No $p$--$\LHe$ correlation function has been measured yet. This section states what the
calculation implies for such a measurement: how the pair can be reconstructed, how momentum
resolution affects the result (Sec.~\ref{sec:exp-res}), which control measurement is needed
(Sec.~\ref{sec:exp-control}), what limits the precision (Sec.~\ref{sec:exp-where}), and how
the calculation should be compared with data (Sec.~\ref{sec:exp-fit}).

\subsection{Reconstructing the pair}
\label{sec:exp-pair}

Every $\LLi$ produced decays into $p+\LHe$: it lives for $\hbar/\Gamma=3.7\times10^{-21}$~s,
and the only other channel below $19$~MeV, $\alpha+p+\Lambda$, opens $2.5$~MeV (fourteen
widths) above the state. Missing-mass spectroscopy~\cite{Bertini1981}, with a resolution of
about $3$~MeV, cannot reach the $\lesssim100$~keV precision needed here, so the correlation
function is the natural observable. The hypernucleus can be reconstructed through its
mesonic decay $\LHe\to\pi^{-}+p+{}^{4}\mathrm{He}$~\cite{Szymanski1991}. STAR already
reconstructs light hypernuclei through their mesonic decays~\cite{STAR2022H34} and measures
baryon correlations, such as $d$--$\Lambda$, in the same collisions~\cite{STAR2025dLambda}.

\subsection{Momentum resolution}
\label{sec:exp-res}

We estimate the effect of momentum resolution by folding the calculated $C(\kstar)$ with
the procedure of Sec.~\ref{sec:exp-fit}, Eq.~(\ref{eq:fold-exp}), modelling the detector
response as an isotropic Gaussian smearing of the vector $\bm{\kstar}$ with width
$\sigma_{\kstar}$ per component. Realistic values of $\sigma_{\kstar}$ follow from the
single-particle resolutions. Propagating fractional momentum errors $\delta p_{i}/p_{i}$ and angular errors
$\delta\theta_{i}$ of the proton and the hypernucleus into $\kstar$, and averaging over the
direction of $\bm{\kstar}$, gives
\begin{equation}
  \sigma_{\kstar}\simeq\frac{\mu\beta}{\sqrt{3}}
  \sqrt{\;\sum_{i}\Bigl(\frac{\delta p}{p}\Bigr)_{i}^{2}
        \;+\;\gamma^{2}\sum_{i}\bigl(\delta\theta\bigr)_{i}^{2}\;} ,
  \label{eq:kres}
\end{equation}
where $\beta$ and $\gamma$ refer to the pair; a Monte Carlo evaluation confirms it to $1\%$.
Because a detector delivers a fractional momentum resolution, the proton and the
hypernucleus contribute comparably despite their different masses. For a momentum resolution
of $0.6$--$1\%$ per particle and pair velocities $\beta=0.5$--$0.8$, Eq.~(\ref{eq:kres})
gives $\sigma_{\kstar}\approx2$--$5\MeVc$. We therefore take $3\MeVc$ as the benchmark and
$5\MeVc$ as a pessimistic case.

\begin{figure*}[tb]
\centering
\includegraphics[width=\textwidth]{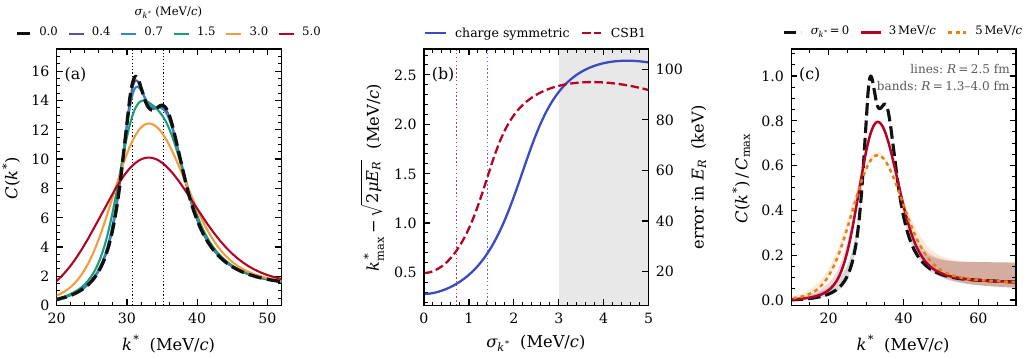}
\caption{Effect of finite momentum resolution on the predicted correlation function, at the
calibrated $\Lambda N$ strength. Fig.~\ref{fig:corr} shows the calculation without
resolution.
(a) The $\LLi$ doublet at $R=2.5$~fm (CSB1), folded with Gaussian resolutions of increasing
width $\sigma_{\kstar}$ [Eq.~(\ref{eq:fold-exp})]; the heavy black dashed curve is the
calculation without resolution. Dotted
vertical lines mark $\sqrt{2\mu E_{R}}$ for the two states. The second maximum disappears
above $\sigma_{\kstar}=0.73\MeVc$.
(b) Displacement of the maximum of the total $C(\kstar)$ from $\sqrt{2\mu E_{R}}$ of the
$1^{-}$ state, against resolution, for CSB1 (dashed) and charge symmetry (solid). The
right-hand axis gives the resulting error in $E_{R}$, to first order in the displacement, if
Eq.~(\ref{eq:kpeak}) is applied to the raw maximum. Dotted vertical lines mark where each
doublet merges; the shaded band is the $3$--$5\MeVc$ range considered in the text.
(c) The same correlation functions, divided by the maximum of the curve without resolution, at
$\sigma_{\kstar}=0$, $3$ and $5\MeVc$. Curves are for $R=2.5$~fm; bands span
$R=1.3$--$4.0$~fm. The recorded shape is nearly independent of the source size, and the
doublet structure of the calculation disappears.}
\label{fig:res}
\end{figure*}

This resolution is not small compared with the structure to be measured: the $1^{-}$ state
is $\mu\Gamma/\kstar_{\rm peak}=4.6\MeVc$ wide in $\kstar$. Figure~\ref{fig:res} shows three
consequences. First, the doublet merges: the second maximum disappears above
$\sigma_{\kstar}=0.73\MeVc$ with CSB1 ($1.42\MeVc$ with charge symmetry), so only the
position of the merged peak can be measured. Second, the maximum moves upward: at $R=2.5$~fm
it shifts from $31.25\MeVc$ at $\sigma_{\kstar}=0$ to $32.34\MeVc$ at $1.5\MeVc$ and
$33.14\MeVc$ at $3\MeVc$, and then saturates ($33.10\MeVc$ at $5\MeVc$). At
$\sigma_{\kstar}=3\MeVc$ the maximum lies $2.4\MeVc$ above $\kstar_{\rm peak}$, so converting
it through Eq.~(\ref{eq:kpeak}) would overestimate $E_{R}$ by about $100$~keV. This shift
comes almost entirely from the unresolved $2^{-}$ state being averaged into the $1^{-}$ peak:
without the $2^{-}$ channel, the maximum moves by less than $0.05\MeVc$ up to
$\sigma_{\kstar}=3\MeVc$. Third, the peak becomes lower (from $44.6$ to $35.3$ at
$R=1.3$~fm, and from $6.0$ to $4.8$ at $R=4$~fm, for $\sigma_{\kstar}=3\MeVc$), and the
folded shape is nearly independent of the source radius.

\subsection{A control measurement: $p$--$\alpha$}
\label{sec:exp-control}

Before the hypernuclear measurement, the analysis should be validated on $p$--$\alpha$
pairs from the same data. The $p$--$\alpha$ system shares every element of the
$p$--$\LHe$ analysis except the $\Lambda$: a proton paired with a $Z=2$ cluster, the same
Coulomb suppression at small $\kstar$, a $P$-wave resonance above threshold, and the same
non-femtoscopic baseline. Unlike $\LLi$, however, the $^{5}$Li $3/2^{-}$ resonance is fixed
by measured $p\alpha$ phase shifts, so its correlation function can be predicted reliably. Our
calculation, with no parameter adjusted on the proton side, places its maximum at
$51.6\MeVc$ for $R=2.5$~fm (grey curve in Fig.~\ref{fig:corr}). Any disagreement therefore
points to the analysis, not to the physics: an error in the momentum scale, in the Coulomb
treatment, in the baseline, or in the source.

The objective is a quantitative agreement. The analysis can be considered under control if
the folded calculation reproduces the measured $p$--$\alpha$ correlation function with a
source radius consistent with that of other light-ion pairs in the same data, and places the
maximum within about $1\MeVc$ of the measured one. This tolerance reflects the model itself:
our $^{5}$Li pole lies $45$~keV below the data-derived pole, which corresponds to
$0.7\MeVc$ in the $p$--$\alpha$ maximum. Agreement at this level bounds a common
momentum-scale error to about $2\%$, which would shift the $p$--$\LHe$ peak by at most
$0.6\MeVc$, well inside the $\pm2.6\MeVc$ uncertainty of the prediction. The control
measurement does \emph{not} test the resolution kernel: the $^{5}$Li resonance is broad
($20\MeVc$ full width in $\kstar$), and its maximum moves only from $51.6$ to $51.7\MeVc$
between $\sigma_{\kstar}=0$ and $3\MeVc$. The kernel for $p$--$\LHe$, a three-track secondary
vertex, must come from embedding and simulation.

\subsection{Statistics and resolution}
\label{sec:exp-where}

The measurement is limited by resolution, not by statistics. Because the peak value of
$C(\kstar)$ is between $6$ and $45$, about $130$ reconstructed pairs within $\pm5\MeVc$ of the
peak at $R=2.5$~fm (about $250$ at $R=4$~fm) determine its position to $0.5\MeVc$, or
$20$~keV in $E_{R}$, assuming a pair purity of $\lambda=0.4$. A limited resolution, in
contrast, increases the systematic uncertainty, because the folded result then depends on
the model's $2^{-}$ state. When the doublet is resolved ($\sigma_{\kstar}\lesssim0.5\MeVc$),
moving the $2^{-}$ to Hiyama's $0.97$~MeV (Sec.~\ref{sec:budget}) changes the measured
maximum by less than $0.1\MeVc$. For $\sigma_{\kstar}=1.5$--$3\MeVc$, the same change shifts
it by $0.8$--$1.2\MeVc$, i.e.\ $30$--$50$~keV in $E_{R}$, more than the statistical error of a
few hundred pairs. Improving the resolution therefore reduces the uncertainty of the result
more than increasing the sample beyond a few hundred pairs.

\subsection{Comparison with the calculation}
\label{sec:exp-fit}

The measured correlation function should be compared with the calculation folded with the
detector response, rather than unfolded. With a few hundred pairs, unfolding is unstable;
more importantly, most of the resolution-induced shift comes from the $2^{-}$ state
(Sec.~\ref{sec:exp-res}), so a correction for it would hide a model dependence inside the
data. Folding keeps that dependence explicit.

The folding acts on pair yields, not on the ratio $C$. Let $M(\kstar,q)$ be the
probability that a pair with true relative momentum $q$ is reconstructed at $\kstar$,
obtained from embedding and simulation, and let $N_{\rm ref}(q)$ be the true
distribution of uncorrelated pairs, obtained from mixed events. The folded correlation
function is then
\begin{equation}
  C_{\rm th}^{\rm folded}(\kstar)
  =\frac{\sum_{q}M(\kstar,q)\,N_{\rm ref}(q)\,C_{\rm th}(q)}
         {\sum_{q}M(\kstar,q)\,N_{\rm ref}(q)} .
  \label{eq:fold-exp}
\end{equation}
Section~\ref{sec:exp-res} uses $N_{\rm ref}\propto q^{2}$ and, for an isotropic
Gaussian smearing of $\bm{\kstar}$ with width $\sigma$ per component,
$M(\kstar,q)=(\kstar/q)\,[g(\kstar-q)-g(\kstar+q)]$, where $g$ is
a normalised Gaussian of width $\sigma$. In an experiment, $M$ and $N_{\rm ref}$ should be
taken from the simulation and the mixed events themselves. The data are then fitted with
\begin{equation}
  C_{\rm exp}(\kstar)=\mathcal{N}\,(1+b\,\kstar)
  \bigl[\,1+\lambda\,(C_{\rm th}^{\rm folded}(\kstar)-1)\,\bigr],
  \label{eq:cexp}
\end{equation}
where $\mathcal{N}$ and $b$ describe the normalisation and the non-femtoscopic baseline, and
$\lambda$ the fraction of genuine primary pairs. The model enters through a set of
calculated curves $C_{\rm th}(\kstar;s,R)$, tabulated on a grid of the $\Lambda N$ strength
$s$ and the source radius $R$. Each curve is folded with Eq.~(\ref{eq:fold-exp}) and fitted
with Eq.~(\ref{eq:cexp}); the source radius can be fixed from other pairs in the same data or
left free. The best-fit $s$ gives $E_{R}$, and hence $\kstar_{\rm peak}=\sqrt{2\mu E_{R}}$,
with $dE_{R}/d\kstar=39$~keV per $\MeVc$. A Poisson likelihood should be used, because few
pairs fall into each bin. Two quantities should be reported: $E_{R}$ (or
$\kstar_{\rm peak}$), and the excess yield
$\int(N_{\rm same}-\mathcal{N}N_{\rm mixed})\,d\kstar$ over $20<\kstar<60\MeVc$, which
constrains $\Gamma$ and, unlike $\int(C-1)\,d\kstar$, is conserved by the folding.

A measured peak position also tests the model itself. Because CSB is degenerate with the
calibration (Sec.~\ref{sec:csb}), a position inconsistent with $30.8\pm2.6\MeVc$ would point
to the frozen core or the two-body reduction, not to CSB.

\section{Limitations}
\label{sec:limits}

Two limitations are reducible and should be distinguished from the rest. First, the
$N\alpha$ potential is fitted to two numbers, the $S$-wave scattering length and the
$^{5}$He resonance energy, with its two ranges fixed by hand; the resulting range
dependence is the $\pm1.1\MeVc$ in the budget. It could be removed by fitting the $N\alpha$
potential directly to the measured $n\alpha$ and $p\alpha$ elastic phase shifts over the
first several MeV, which are among the best-determined scattering observables in light
nuclei, instead of to derived resonance parameters. Second, converting the emulsion
$B_\Lambda(\LHesix)$ into the neutron separation energy used in the calibration requires a
$^{5}$He energy, and the choice between the evaluated pole and the mass-table value is worth
$1.3\MeVc$. Establishing which $^{5}$He mass underlies the compiled $B_\Lambda$, or quoting
$B_{n}$ directly from measured masses, would remove this uncertainty. We have done neither
here; these are the model-side changes most likely to sharpen the prediction.

The model omits several other effects that a quantitative analysis would eventually need.
Breakup into $p+\alpha+\Lambda$ opens at a relative energy of $3.12$~MeV, i.e.\ near
$\kstar\approx70\MeVc$, and is not included; this lies above the resonance region but
affects the tail. Core polarisation is absent by construction, and $\Lambda NN$ three-body
forces enter only through the empirical factor $s$. The non-$\LHe$ admixture and the core
contraction found in Ref.~\cite{Hiyama1996} (Sec.~\ref{sec:fold}) are of the same order as
the $7$--$9\%$ excess in our halo radii (Sec.~\ref{sec:A6}). This accounting is not exact, because our frozen
$\alpha$--$\Lambda$ distance of $3.15$~fm is itself $13\%$ larger than the $2.79$~fm that
Ref.~\cite{Hiyama1996} obtains for free $\LHe$ at the same $B_\Lambda$; the two calculations
therefore also differ in the $\Lambda\alpha$ potential. The ordering of the $\LLi$ doublet is
an input inherited from the chiral $\Lambda N$ interaction (Sec.~\ref{sec:pw}). The predicted
position of the merged peak is an intensity-weighted average of the two states and therefore
depends on this assumption, which a measurement cannot test because it cannot resolve the
two peaks.
The proton spin--orbit interaction is folded only approximately, and $\LHe$ is treated as
point-like in the emission source, whereas its finite size should in principle be folded
into $S(r)$. No feed-down or $\lambda$-parameter dilution is applied to the theoretical
curves.

Two further caveats deserve emphasis. First, the $S$-wave $p\alpha$ interaction is
represented by a deep attractive potential with a Pauli-forbidden state. The
phase-equivalent shallow repulsive alternative gives identical $f_{0}$ and $d_{0}$ but a
different interior wave function, and therefore a different $C(\kstar)$ for small sources.
This is another reason why the $S$-wave scattering parameters do not determine the
correlation function in this system. It is also a concrete instance of a general problem
identified by Epelbaum \textit{et al.}~\cite{Epelbaum2025}: a source function cannot be
assumed universal across interaction models, but should be off-shell consistent with the
interaction it is folded with. We use the same Gaussian source for all interaction
variants, so any difference between interaction variants, in particular between the two
phase-equivalent $p\alpha$ representations, would carry this model dependence. Our central
observable, the peak \emph{position}, is fixed by the $S$-matrix pole rather than by the
short-distance wave function, and we expect it to be far more robust in this respect than
the peak height. Smith and Blum~\cite{Smith2026} have quantified related corrections to the
smoothness and on-shell approximations and find them at or below the percent level for $pp$
correlations and deuteron coalescence. Second, the CSB analysis rests on a single scenario.
With older $A=4$ data the pattern changes, and our mapping of CSB onto Gaussian depths is
cruder than the underlying contact-plus-one-pion-exchange structure. As Sec.~\ref{sec:csb}
shows, the resulting shift is comparable to the residual model dependence of the $\LLi$
pole, so we treat it as a variation of the prediction, not as an effect to be tested.

Finally, the predicted $\LLi$ $1^{-}$ state has never been observed directly. The only
empirical information is a single $(K^{-},\pi^{-})$ measurement of
$B_\Lambda(\LLi)$~\cite{Bertini1981}. It rests on a weak transition of uncertain assignment,
has an uncertainty of $\pm0.5$~MeV (Sec.~\ref{sec:A6}), and gives no information on the
width. The only previous continuum calculation~\cite{Oishi2018} obtains a narrower state,
with a width of $0.01$--$0.1$~MeV. A correlation measurement would address both the position
and the width, and it is at present the only route to the width.

\section{Summary}
\label{sec:summary}

We have computed the $p$--$\LHe$ femtoscopic correlation function in an $\alpha+\Lambda$
folding model. The two-body input is fixed by $n\alpha$ scattering, the $^{5}$He resonance,
$B_\Lambda(\LHe)$ and the chiral NLO $\Lambda N$ scattering parameters. A single overall
$\Lambda N$ strength factor is calibrated on the measured $\LHesix$ binding energy:
$s=0.891^{+0.072}_{-0.076}$ in the CSB1 scenario adopted throughout ($0.915$ with charge
symmetry). Without any fit on the proton side, the model reproduces the $^{5}$Li $3/2^{-}$
resonance, whose pole lies $45$~keV below the data-derived pole; the broad $1/2^{-}$ is
placed $0.44$~MeV low. The model predicts the mirror $\LLi$ $1^{-}$ resonance at
$E_{R}=0.602$~MeV with $\Gamma=0.178$~MeV ($0.553$ and $0.149$~MeV with charge symmetry).

The correlation function has a large peak at $\kstar\simeq31\MeVc$. The maximum of the
resonant $1^{-}$ channel satisfies $\kstar_{\rm peak}=\sqrt{2\mu E_{R}}$ to within $1\%$
across the calibrated band ($3\%$ without the $\Lambda N$ interaction), and the unresolved
$2^{-}$ state shifts the maximum of the total $C(\kstar)$ up by a further $0.26\MeVc$. An
exact partial-wave decomposition shows that the $P$ wave carries the entire enhancement over
the Coulomb baseline, $79\%$ of it from the $1^{-}$ channel; the $S$ wave contributes a small
negative term, and the $D$ and $F$ waves are negligible at the peak, though not above it.
The Coulomb-modified $S$-wave parameters are well defined ($f_{0}=-3.67$ and $-4.08$~fm,
$d_{0}=+0.60$ and $+0.67$~fm for singlet and triplet), but they carry almost none of the
measurable signal, and an $S$-wave Lednick\'y--Lyuboshitz analysis is inapplicable to this
system.

Our treatment is simpler than the established three-body and \emph{ab initio} descriptions
of the $A=6$ hypernuclei and does not compete with them on structure. Instead, it provides
what those calculations do not: the femtoscopic observable, resolved channel by channel,
with its dependence on source size and momentum resolution, and the effect of
charge-symmetry breaking. As hypernucleus--nucleon correlations come within experimental
reach, theoretical predictions will be needed in this form.

Because the measurement has not yet been made, we have also described what it requires
(Sec.~\ref{sec:exp}). Reconstructing $\LHe$ and measuring baryon correlations are already
established, for example at STAR, and every $\LLi$ produced appears as a $p$--$\LHe$ pair.
The demanding requirement is the momentum resolution. A $\kstar$ resolution near $3\MeVc$,
two thirds of the intrinsic width $\mu\Gamma/\kstar=4.6\MeVc$, is needed, and the proton and
the hypernucleus contribute comparably to it. At this resolution the measured maximum lies
$2.4\MeVc$ above $\kstar_{\rm peak}$, almost entirely because the unresolved $2^{-}$ state is averaged
into the $1^{-}$ peak; the calculation must therefore be folded with the resolution, not the
data unfolded. A $p$--$\alpha$ measurement on the $^{5}$Li $3/2^{-}$ resonance should come
first: it tests the momentum scale, the non-femtoscopic baseline and the Coulomb treatment,
though not the resolution kernel. Because the peak value of $C(\kstar)$ is between $6$ and
$45$, a few hundred pairs determine the peak position to $0.5\MeVc$, or $20$~keV in $E_{R}$.
The precision is therefore limited by resolution, not by statistics: a poorer resolution
makes the result more dependent on the unconstrained $2^{-}$ state and so increases the
systematic uncertainty. The quantities to report are $\kstar_{\rm peak}=\sqrt{2\mu E_{R}}$
from a forward-folded fit and the integrated excess over $20<\kstar<60\MeVc$.

\begin{acknowledgments}
We are grateful to Emiko Hiyama and Lisheng Geng for valuable discussions and constructive suggestions on this work. This work was supported by the National Natural Science Foundation of China with Grant No.~12375141 and 124B200053.
\end{acknowledgments}


\end{document}